\documentclass[twocolumn]{revtex4-1}

\usepackage{amsmath, amsfonts, amssymb, graphicx, xcolor}
\usepackage[normalem]{ulem}
\graphicspath{{./figs/}}
\usepackage{multirow}
\usepackage{makecell}

\newcommand{\siref}[1]{Appendix~\ref{#1}}

\newcommand{\etal}{\textit{et al.}}
\newcommand{\probP}{\text{I\kern-0.15em P}}
\newcommand{\AW}{}
\newcommand{\TM}{}
\newcommand{\DW}[1]{{#1}}
\newcommand{\DWstrike}[1]{{\color{red}\sout{#1}}}
\renewcommand{\sout}[1]{}

\newcommand{\equal}{Corresponding authors. These authors contributed equally.}
\newcommand{\ENS}{{Laboratoire de physique de l'\'Ecole normale sup\'erieure,
		CNRS, PSL University, Sorbonne Universit\'e, and Universit\'e 
		Paris-Cit\'e, 75005 Paris, France}}

\begin{document}
\affiliation{\ENS}
\author{Daniel PGH Wong}
\affiliation{\ENS}
\author{Aleksandra M. Walczak}
\thanks{\equal}
\affiliation{\ENS}
\author{Thierry Mora}
\thanks{\equal}
\affiliation{\ENS}


\newcommand{\deftitle}{{\TM Statistical survey of the adaptive landscapes of 10,000 antibodies}}

\title{\deftitle}

\begin{abstract}
Affinity maturation is the Darwinian process by which antibodies improve antigen binding through somatic hypermutation and selection. The adaptive landscape, which defines the set of antibody-specific mutations that \DW{improve \sout{antigen binding or other} functional characteristics}, has been explored in only a handful of antibodies. Identifying the sites of adaptive mutations in a given sequence, and how these sites vary across the antibody repertoire, can inform the design of therapeutic antibodies. We develop a \DWstrike{parameter-free} population genetic framework that leverages the statistics of convergent affinity maturation in B cell lineages sharing similar naive sequences, called public clonotypes, to identify beneficial mutations. Applying this framework to more than 10,000 public clonotypes represented by multiple lineages across 20 healthy individuals, we identify widespread signatures of \DW{\sout{clonotype-dependent selection of individual mutations}site-specific selection. We estimate characteristic fixation rates of beneficial} mutations across the V gene\DW{\sout{ at the single-site level}}, \DW{distinguishing mutations that are rarely but strongly beneficial across clonotypes, from those that are commonly but only weakly beneficial}. The \DW{statistics of these} inferred landscapes broadly reproduce the \DW{\sout{statistics} rates} of convergent mutation in antibodies specific to SARS-CoV-2 and influenza. Finally, we use our framework to benchmark predictions from existing antibody language models, and show that while these models are dominated by non-selective signatures, a simple renormalization procedure can expose signatures of clonotype-dependent positive selection consistent with our predictions.

\end{abstract}

\maketitle

\section{Introduction}

During affinity maturation, the pathogen-binding receptors of selected B cells undergo mutation and selection to improve binding to their cognate antigen. 
Much work has gone into characterizing the molecular and cellular mechanisms of affinity maturation, and how they combine to reliably improve the average affinity of a population of antigen-specific B cells~\citep{maclennan1994germinal, victora2012germinal, victora2022germinal}.
However, the evolutionary fate of any individual B cell lineage undergoing 
affinity maturation remains harder to predict. This intrinsic uncertainty is due to the stochastic nature of evolutionary forces like somatic hypermutation (SHM) and genetic drift, as well as the small number of mutations available to a given antibody that improve affinity to its cognate antigen~\citep{dewitt2025replaying}. The identities and effects of these mutation depend on the ancestral naive rearrangement {of the receptor sequence}, which defines the {B cell's} clonotype. Other mutations may also be selected for improved folding stability and expression, or reducing self-reactivity. The overall impact of such mutations for B cell survival during affinity maturation defines each clonotype's ``adaptive landscape." Measuring these landscapes is needed to understand the consistency and predictability of affinity maturation of a given B cell lineage. For instance, how much more likely is a typical beneficial mutation to fix than a neutral mutation with an equivalent mutation rate? How likely are two independent lineages, founded with the same naive sequence, to follow correlated mutational trajectories? More practically, a global survey of adaptive landscapes could guide predictions of adaptive mutations in naturally derived or \textit{in silico} designed antibodies of interest. 

High-throughput experimental approaches like deep mutational scans, as well as exhaustive combinatorial libraries of up to 16 distinct mutations, have been used to measure the binding properties of large numbers of antibody-antigen pairs and provide insight into the local structure of the binding landscape~\citep{koenig2015deep, adams2016measuring, koenig2017mutational, madan2021mutational, dewitt2025replaying, phillips2021binding, moulana2022compensatory, moulana2023landscape, schulz2025epistatic, kirby2025retrospective}. However, each antibody necessitates the construction of its own variant library, which must then be measured against each individual antigen. As a result, only a handful of landscapes, representing a tiny fraction of antibody and antigen diversity, have been measured experimentally. Additionally, how these \textit{in vitro} measurements map to \textit{in vivo} fitness must be separately determined.

Repertoire sequencing provides an alternative, scalable view of affinity maturation evolutionary landscapes, by directly sampling the outcomes of thousands or millions of evolutionary processes through the B cell lineages that survive in an individual. Unlike exhaustive experimental mapping, only narrow parts of the landscape are ever visited via SHM in a given lineage, and then sampled by sequencing. Thus, repertoire sequencing trades off deep coverage of a single landscape for sparse coverage of thousands or millions of landscapes at once. The effect of the underlying evolutionary landscape can be identified by mutational signatures of positive selection in each lineage, but these signatures must be distinguished from the effects of other evolutionary forces, namely SHM biases, neutral drift, and genetic hitchhiking~\citep{nourmohammad2019fierce}. Previous studies have identified positive selection by studying population genetic statistics over many sequences, typically variants of dN/dS, which measures the ratio of nonsynoymous to synonymous mutations, measured along an affinity matured lineage's phylogenetic tree~\citep{yaari2012quantifying, yaari2015mutation, horns2019signatures, nourmohammad2019fierce, hoehn2021human, ralph2020using, mikelov2022memory}. Because these approaches rely on aggregating mutation signal across an antibody sequence, or averaging signal across many clonotypes, they cannot provide site-level resolution of the clonotype-dependent adaptive landscape for different receptor sequences.

Repertoire data has also been used to train antibody-specific protein language models (ALM), with the goal of using the high-dimensional expressivity of deep neural networks to encode the sequence dependent features of selection landscapes inferred from mutation patterns across millions of sequences~\citep{ruffolo2021deciphering, olsen2022ablang, shuai2023iglm, kenlay2024large, burbach2024improving, burbach2025curriculum,lu2026conditionally}. \DW{These models, and more generally protein language models, hold promise in both protein design and optimization, as well as in generative modeling of protein and repertoire evolution. However, existing ALMs} mostly learned non-selective signatures, like biases in the statistics of VDJ rearrangements or the nucleotide sequence context bias of SHM. While recent ALMs control for these biases~\citep{olsen2024addressing, ng2025focused, matsen2025sitewise, matsen2026separating,lu2026conditionally}, it is not clear that they learn features of the adaptive landscape that are specific to each clonotype\DWstrike{ and its cognate antigen}. Indeed, it is unknown whether current repertoire data, lacking information about the corresponding antigens, contains sufficient signal for any inference method---language model or otherwise---to learn the patterns of clonotype-dependent, let alone antigen-dependent, beneficial mutations. Without addressing this question, it is unclear if current failure modes in ALMs reflect deficiencies in architecture, training recipes, or in the data itself.

We develop a repertoire-based approach to study the adaptive landscape of affinity maturation, focusing on tens of thousands of ``public" clonotypes represented by independent and affinity matured lineages in multiple individuals. These lineages of shared clonotypes \DW{are thought to typically} reflect convergent response to the same, although \DW{\sout{generally} sometimes} unknown, antigen~\citep{rao2025public}. Such public antigen-specific clonotypes have been observed in individuals infected with HIV~\citep{setliff2018multi, murji2022sequence}, HCV~\citep{skinner2023convergent}, Ebola~\citep{ehrhardt2019polyclonal}, influenza~\citep{pappas2014rapid,forgacs2021convergent, wang2024explainable}, and SARS-CoV-2~\citep{robbiani2020convergent, galson2020deep, chen2021convergent, tan2021sequence, wang2022large}. In some cases, convergent mutations have been validated as affinity enhancing mutations~\citep{tian2022prominent, tan2021sequence, wang2022large, rao2025clonotype, niu2025ai, pappas2014rapid}.  We analyze these shared-clonotype lineages within a population genetic framework to study the statistics of convergent evolution across thousands of shared clonotypes that are detected in the bulk repertoires of healthy individuals.

\section{Results}
\subsection{Twenty thousand expanded lineages with shared clonotypes across subjects}

\begin{figure*}
    \centering
    \includegraphics[width=0.95\textwidth]{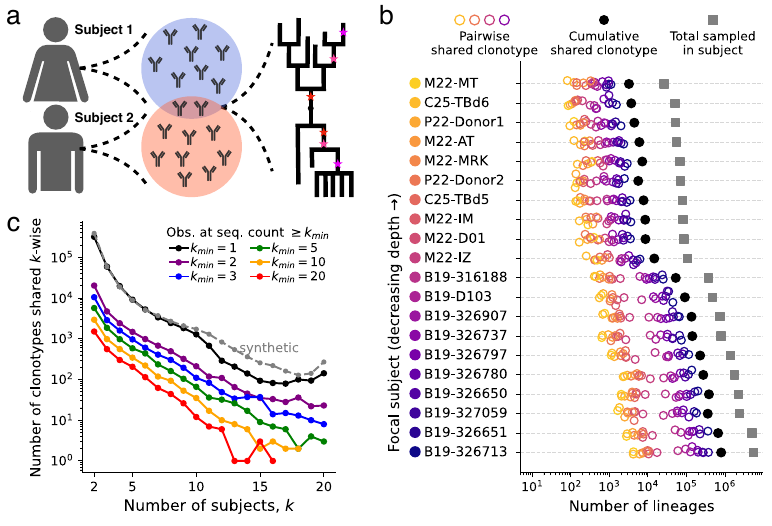}
    \caption{\textbf{Thousands of shared clonotypes uncovered by deep repertoire sequencing}. (\textbf{a}) Lineages of shared 
      clonotypes, defined as lineages sharing the same V and J genes and CDR3 lengths, and CDR3 amino acid identity {${>}85\%$}, are identified among B cell repertoires deriving from different individuals.
    (\textbf{b}) For each focal subject on the y axis: number of lineages sharing a clonotype with the 19 other subjects across four studies: \textbf{B19}~\citep{briney2019commonality}, \textbf{P22}~\citep{phad2022clonal}, \textbf{M22}~\citep{mikelov2022memory}, \textbf{C25}~\citep{cvijovic2025long} (open circles, color-coded by subject). Black circles: total number of those lineages sharing a clonotype with at least one other subject. Grey squares: total number of lineages {sampled in each} subject. (\textbf{c}) Number of clonotypes shared by $k$ subjects, for different minimum lineage sizes (colors). The number of shared clonotypes with no lineage size restriction (black curve) is consistent with the prediction from the SoNNia computational model of recombination and global selection (dashed grey curve, {see Methods}).}
    \label{fig1:schematic}
\end{figure*}

We leveraged antibody heavy-chain repertoire data across 20 healthy subjects, drawn from two bulk VDJ RNA sequencing studies by Briney~\etal~(henceforth \textbf{B19}~\citep{briney2019commonality})
and Mikelov~\etal~(\textbf{M22}; \citep{mikelov2022memory}); and two single cell VDJ sequencing studies by Phad~\etal~(\textbf{P22};~\citep{phad2022clonal}) and Cvijovic~\etal~(\textbf{C25};~\citep{cvijovic2025long}). These studies were chosen for their substantial per-subject depth and Unique Molecular Identifier (UMI)-based amplification protocols to correct for PCR and sequencing errors (Methods and SI \siref{si:datasets}). The 20 sequenced subjects were healthy adults
living in California, Western Europe, or Russia at the time of sampling. Thus, any shared sequences identified between subjects likely reflect related pathogen exposures over subjects' lifetimes, such as vaccination or pathogen challenges. \DWstrike{This focus on healthy people stands in contrast to systematic studies of sequence sharing across repertoires in subjects under acute or chronic infection or vaccination~\citep{setliff2018multi}.}

We first processed and sorted each subject's repertoire into memory lineages (Methods and \siref{si:sequence processing}). In brief, \DW{after standard application of antibody alignment and germline annotation methods,} we sorted each subject's \DW{affinity matured or} ``memory" sequences\DW{, containing ${\geq}3$ mutations in their templated (i.e. alignable) regions,}  into their respective VJ$\ell$ classes, defined by shared Variable (V) and Joining (J) gene annotations and length of the complementarity determining region (CDR) 3. Within each VJ$\ell$ class, we then \DWstrike{applied HILARy~\citep{spisak2024combining}, which leverages both CDR3 distance and mutation patterns in templated (non-CDR3) regions, to} clustered sequences into lineages sharing the same naive ancestor {\TM(Methods)}. After processing, we observed similar statistics of V and J gene usage, CDR3 length, and lineage size across subjects and studies, despite differences in anatomical sampling sites, cell-type sorting, and sequencing modalities (SI Fig.~\ref{sfig:subject repertoire statistics}).

We next studied the statistics of B cell memory clonotype sharing between subjects. We defined two lineages as sharing a clonotype between two individuals if they shared the same V and J genes, same CDR3 lengths, and CDR3 amino acid identity larger than $85\%$ (Fig.~\ref{fig1:schematic}a), consistent with previous definitions of inter-subject ``public" clonotypes~\citep{setliff2018multi, wang2022large, rao2025clonotype}. We identified around 3 million memory lineages, representing 15\% of all lineages, whose clonotypes were represented in two or more subjects~(Fig.~\ref{fig1:schematic}b).  Sharing rates were comparable between pairs from different studies after adjusting for subject-specific sampling depth (SI Fig.~\ref{sfig:sharing vs pairwise comparisons}a), suggesting that contamination played a negligible role in the observed sharing.
In addition, more than $80\%$ of shared-clonotype lineages across individuals from the same study differed in at least one position in their consensus sequence (defined at each site as the nucleotide found in more than $80\%$ of lineage's sequences, see SI \siref{si:sequence processing} and Fig.~\ref{sfig:sharing vs pairwise comparisons}b), further ruling out contamination as a substantial contribution. 

By combining many subjects from multiple studies, the number of shared clonotypes we identified is at least an order of magnitude larger than reported
in previous work. \DWstrike{MOVE TO SI? For instance, P22~\citep{phad2022clonal} identified 138 shared memory clonotypes between their two donors. By our analysis, each of these donors harbors more than $4000$ memory lineages that share clonotypes with at least one other subject from other studies.} A typical B19 donor harbors more than $100,000$ lineages containing ``public" clonotypes (Fig.~\ref{fig1:schematic}b). Since the expected number of sharing clonotypes grows with the number of pairwise comparisons, it strongly depends on sequencing depth \citep{ruiz2023modeling}.
The rates of pairwise and higher-order sharing (Fig.~\ref{fig1:schematic}c) were quantitatively consistent with simulated repertoires generated from a computational null model of VDJ recombination and \DW{\sout{global} antigen-agnostic} selection~\citep{elhanati2015inferring,marcou2018high,sethna2019olga,isacchini2021deep}, trained on predominantly naive-like sequences (fewer than $3$ mutations from germline, see Methods). \DW{Shared clonotypes were representative of the underlying repertoire diversity in each subject. While shorter CDR3s are more likely to be shared, the V and J gene usage of  shared clonotypes at fixed CDR3 length were quantitatively representative of each subject's background distribution at that length~(SI Fig.~\ref{sfig:vj3 shared usage}).}

Of particular interest was clonotype sharing among \textit{expanded} lineages, represented by at least two sequences in a given subject (Fig.~\ref{fig1:schematic}c). Expanded lineages carry more signal about the evolutionary processes operating at different sites along the sequence, since it is possible to distinguish high-frequency from low-frequency mutations~\citep{yaari2015mutation}. We identified more than 20,000 memory clonotypes coming from expanded lineages in two or more individuals (purple curve in Fig.~\ref{fig1:schematic}c).
\DW{We study these shared expanded lineages as the outcomes of a natural and massively parallel replicate evolution experiment.}

\subsection{Convergent fixation of mutations in distinct lineages of shared clonotype}  
\begin{figure*}
    \centering
    \includegraphics[width=1.0\linewidth]{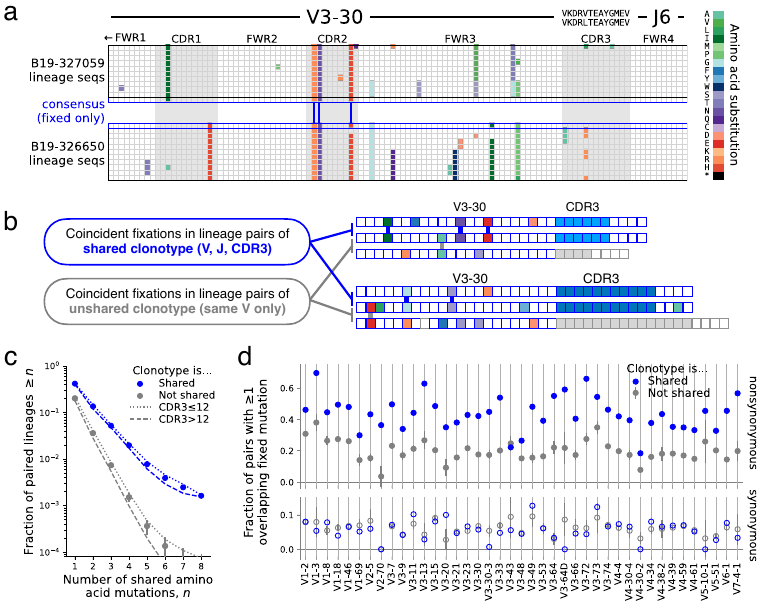}
    \caption{\textbf{Coincident mutations in lineages with a shared clonotype.} 
      (\textbf{a}) Example amino acid alignment of lineages from two subjects sharing the same clonotype. Each lineage contains 10 sequences.
      White (FWR regions) and grey (CDR regions) backgrounds indicate naive (germline) ancestor. Amino acid changes are shown in color
      Each lineage is summarized by a consensus sequence of fixed substitutions (bordered in blue), defined as measured in at least $80\%$ of sequences. Coincident fixations between the two lineages are connected by blue bars. 
    (\textbf{b}) Schematic for the empirical null model. The number of coincident fixed mutations in the V gene between shared-clonotype lineage pairs are compared to those in lineage pairs of unshared clonotype, but sharing the same V gene and a similar number of fixed amino acid mutations. 
    (\textbf{c}) Convergent cumulative distribution of the number of coincident fixed substitutions in pairs of lineages with a shared clonotype (blue), versus pairs of lineages with unshared clonotype (grey), restricted to lineages with at least two unique sequences and at least one fixed amino-acid mutation. 
    (\textbf{d}) Fraction of shared-clonotype (blue) and null (gray) lineage pairs with at least one coincident substitution, stratified by V genes supported by 10 or more shared clonotype pairs. The top panel shows fixed amino acid substitutions, while the bottom panel shows fixed synonymous mutations.
    }
\label{fig2:correlations}
\end{figure*}

\DW{We first asked whether these evolutionary replicates exhibited correlations in their basic properties, such as lineage abundance (SI Fig.~\ref{sfig:abundance correlations}) and number of mutations, a proxy for time spent in germinal centers (SI Figs.~\ref{sfig:macroevo correlations},~\ref{sfig:gene specific macroevo correlations}). We found significant, if modest, correlations in both. This result indicates that, despite differences in individual exposure histories and the stochasticity of clonal dynamics, convergent parallel evolution of clonotypes across individuals is still detectable.}

We next examined whether particular mutations at specific sites were convergent in
shared-clonotype expanded lineages from different subjects. Fig.~\ref{fig2:correlations}a shows a protein sequence alignment of sequences belonging to two lineages in two different individuals that differ by only one amino acid in their CDR3; additional examples are shown in SI Fig.~\ref{sfig:example amino acid alignments}. Many amino acid substitutions are private to a single individual's lineage, but many are also shared, which we call ``coincident fixations." For each lineage, we define as fixed the mutations that are observed in at least $80\%$ of sequences belonging to that lineage. \DW{For lineages represented by just two sequences, which is the majority of lineages, this condition implies observation of the mutation in both sequences. This does not allow for a precise estimate of its frequency, but is enough to ensure the mutation is represented by a very large number of cells and has reached a large fraction of the lineage~(\siref{si: pop gen framework}).}

To systematically quantify the level of site-specific convergence, for each pair of shared-clonotype lineages (each composed of two or more sequences), we identified the sites \DW{in the V gene} where the mutations fixed to the same nucleotide or amino acid in the two lineages. \DW{We restrict to mutations outside of the CDR3 because such mutations, while important, are difficult to unambiguously impute since the ancestral sequence is generally unknown.} {\AW Mutations in the light chain can also contribute to affinity maturation [46], but the bulk heavy-chain data that make up most of our dataset do not pair heavy and light chains, putting light-chain mutations outside the scope of this analysis.} 
We compared the statistics of these coincident mutations among around $10,000$ pairs of shared-clonotype lineages versus an empirical null expectation constructed by ``unpairing clonotypes,'' i.e. by swapping one lineage in each pair with another random one from the same subject
(Fig.~\ref{fig2:correlations}b).
To prevent possible confounding factors the null lineage was chosen to have the same V gene and \DW{\sout{the same} at least as many} fixed amino acid mutations,
but a randomly paired CDR3 and thus a different clonotype \DW{(Methods)}. This constrained swap preserves the V-specific SHM landscape within null pairs, but eliminates any selection-driven correlations relating to shared CDR3 or J gene. 
Across the shared lineages, we observed a substantial enrichment of lineages with shared amino acid substitutions relative to the null (Fig.~\ref{fig2:correlations}c; SI Fig.~\ref{sfig:vfam coincidence survival functions} for enrichments restricted to lineage pairs from different datasets). This convergence in mutational fate is consistent for short and long CDR3 lengths, as well as V family~(SI Fig.~\ref{sfig:vfam coincidence survival functions}), showing that these evolutionary correlations persist across dimensions of B cell receptor diversity.

{At the level of specific V genes}, we found that $40$-$60\%$ of shared-clonotype lineage pairs share at least one fixed amino acid change in the V gene---two to three times more frequent than the null expectation (Fig.~\ref{fig2:correlations}d). This result persists when only considering lineage pairs deriving from different datasets, again ruling out intra-dataset contamination (SI Fig.~\ref{sfig:vgene panel different datasets}a). Importantly, enrichment is absent when considering synonymous mutations (Fig.~\ref{fig2:correlations}d). This rules out explanations for coincident enrichment whereby some antigens would drive stronger germinal center reactions than others, leading to more fixed mutations {\TM and thus more coincidences (including synonymous ones)}. {\TM\sout{This would result in an apparent enrichment of shared fixed mutations relative to the null, but one that would also be observed for synonymous mutations. Since we do not observe this enrichment for synonymous mutations,}} We conclude that the observed enrichment for non-synonymous mutations reflects \DW{clonotype-dependent selection for specific mutations}.
We repeated the analysis while including non-expanded lineages represented by a unique sequence.
While increased sharing of nonsynonymous mutations among shared-clonotype lineages is still observable (Fig.~\ref{sfig:vgene panel different datasets}b), the signal is much weaker. This emphasizes how focusing on high-frequency mutations in expanded lineages acts as a more stringent filter of clonotype-dependent selection than mere co-occurrence.

\DW{We also studied how the observed enrichment depends on the CDR3 similarity threshold used to define shared clonotypes~(SI Fig.~\ref{sfig:cdr3 distance threshold analysis}). A more liberal threshold identifies more shared clonotypes, but which are more likely to respond to different antigens. This is reflected in a reduced enrichment of coincident mutations relative to the null (SI Fig.~\ref{sfig:cdr3 distance threshold analysis}c,d), which disappears completely for similarity thresholds of $70\%$ and below. Our intermediate $85\%$ threshold balances two competing interests: identifying the largest possible number of shared clonotypes, while ensuring the functional similarity and consequent evolutionary correlations of the representative lineages.}

\subsection{Convergent mutations are enriched in and near complementarity determining regions 1 and 2}

\begin{figure*}[t!]
    \centering
    \includegraphics[width=1.0\linewidth]{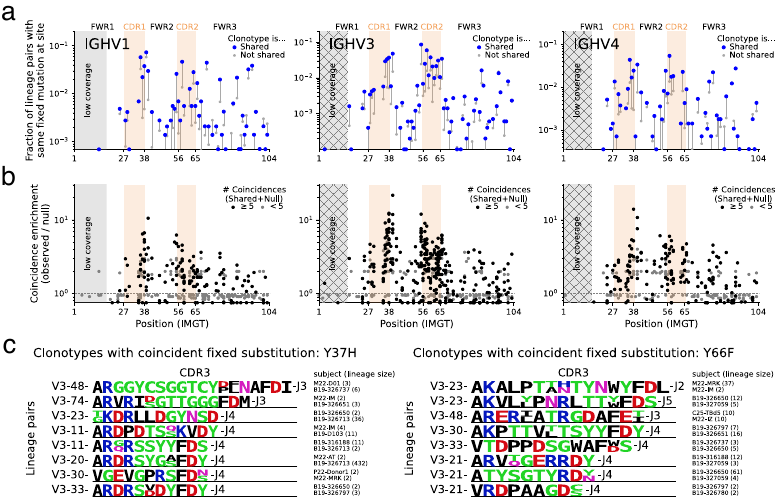}
    \caption{\textbf{Landscape of coincident mutations in shared-clonotype lineages}. (\textbf{a}) Fraction of lineage pairs sharing the same fixed amino acid substitution at each IMGT-aligned site, summed over mutations at each site, for the 3 most abundant V gene families. Lineage pairs with shared clonotype (blue) show consistently higher rates than lineage pairs with unshared clonotypes (grey). Sites with a single coincident mutation lie on the x-axis line. (\textbf{b}) Ratio of shared- versus unshared- clonotype coincidence rates for fixed amino-acid mutation (ratio of blue to grey dots in \textbf{a}). Black versus grey points indicate repeatedly (5 and more) and sparsely (less than 5) sampled coincident mutations, respectively. The vast majority of mutations have enrichments larger than 1; for clarity, the handful of enrichments below 1 are bounded from below on the x-axis line. (\textbf{c}) Clonotype logo plots of V3-family lineage pairs with coincident fixation (${\geq}80\%$ frequency) of mutations Y37H (left) and Y66F (right). Eight pairs, among 123 (Y37H) and 68 (Y66F), were chosen for illustration. Each lineage in the pair is represented by its consensus CDR3 in the logo plot: sites with distinct residues in the consensuses of the two lineages are indicated by two equally sized stacked letters. }
    \label{fig3:landscapes}
\end{figure*}

To determine which sites contributed to selection acting on specific clonotypes, we calculated the fraction of shared-clonotype lineage pairs that had a coincident amino acid mutation in the V gene, both in the data and in the empirical null~(Fig.~\ref{fig3:landscapes}a). The observed rates of site-level mutational amino acid coincidence across lineage pairs ranged from below $0.1\%$ to as high as $10\%$. Some of these coincidences can be explained by strong local sequence biases of the SHM machinery, as can be \DW{\sout{quantified}bounded} by the empirical null based on pairs of lineages with different clonotypes but the same V gene (grey points), which share the same SHM targeting bias.
However, the coincidence rate is higher in pairs of shared-clonotype lineages (blue points), consistent with  Fig.~\ref{fig2:correlations}c,d.
This enrichment can be summarized by the ratio of coincident mutations in pairs of lineages with a shared clonotype to pairs of lineages with random unshared clonotypes (Fig.~\ref{fig3:landscapes}b). This per-site ratio accounts for site-to-site SHM rate variation, and can be interpreted as a measure of the strength of clonotype-dependent selection at each site.
Much of this enrichment is concentrated around the border between the CDR1 and the framework region (FWR) 2, and in the CDR2 region, as expected since these regions are more likely to contribute to binding and less constrained by folding stability than the framework regions~\citep{chothia1989conformations}.
  
To visualize what clonotypes are associated with particular mutations, Fig.~\ref{fig3:landscapes}c presents sequence logos of the CDR3 of 16 representative clonotypes whose paired lineages exhibited the two representative convergent mutations Y37H (123 clonotypes versus 17 in the null) and Y66F (68 clonotypes versus 19 in the null); see SI Fig.~\ref{sfig:more logo plots} for more examples of mutations.
The represented clonotypes exhibit a diversity of CDR3s and come from different subjects. These examples show that our enrichment statistic uncovers rich patterns of convergently selected mutations.

\subsection{Disentangling the \DW{frequencies and fixation probabilities} of clonotype-dependent beneficial mutations} 

\begin{figure*}
    \centering
    \includegraphics[width=1.0\linewidth]{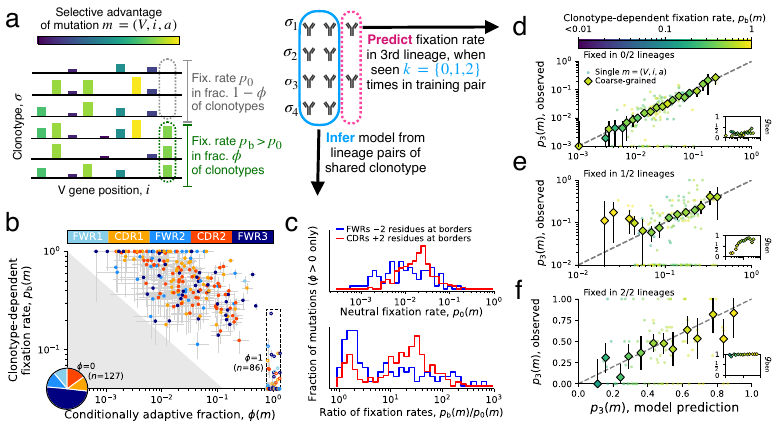}
    \caption{\textbf{Inferring how often each mutation is beneficial across clonotypes, and how much}. 
      (\textbf{a}) Minimal model: for each possible mutation $m=(V,i,a)$ (graphically represented by its site $i$ for simplicity), we define the fraction $1-\phi(m)$ of clonotypes in which that mutation is neutral (top 3 clonotypes for the example mutation outlined by a dotted line), with fixation probability $p_0(m)$, and the fraction $\phi(m)$ in which it is beneficial (bottom 3), with fixation rate $p_{\rm b}(m)$. The selective advantage $p_{\rm b}(m)/p_0(m)>1$ is represented by colored bars. 
      (\textbf{b}) Inferred $\phi(m)$ and $p_{\rm b}(m)$ for each mutation $m$, color-coded by its site's V gene region, \DW{for 426 mutations that were sufficiently well-measured and consistent with the assumptions of the minimal model. {\TM The grey region denotes infeasible parameter space, reflecting the constraint of finite data.
      127 mutations had coincident fixation rates consistent with the neutral expectation, and were set to $\phi= 0$ (pie chart at bottom left). Another 86 mutations' fixation rates were consistent with being always selective, $\phi=1$. Error bars show the interquartile range obtained by bootstrap. \DW{For clarity, only mutations with at least $2$ instances of coincident fixation among paired lineages are shown.}}}
      (\textbf{c}) Distribution of neutral fixation probability $p_{0}$ (top) and selective advantage $p_{\rm b}/p_0$ (bottom) across mutation in the CDRs (extended by 2 amino acid positions on each side) and FWRs (trimmed by 2 amino acid positions on each end). 
      (\textbf{d-f}) Model validation on held-out lineage. For each mutation $m$, model-predicted vs observed frequency $p_3$ of clonotypes in which $m$ will fix in a third lineage conditioned on its fate in the first two lineages that share the same clonotype: fixed in (\textbf{d}) 0 (\textbf{e}) 1 or (\textbf{f}) 2 out of the first two lineages. Each small point is a single $m$ colored by $p_{\rm b}(m)$. Diamonds give averages over bins along the x axis of log-width = 0.1.
      Insets: relative contribution $g_{\rm ben}$ to $p_3$ from clonotypes in which $m$ is beneficial (see main text).
    }
    \label{fig4:mixture model inference}
\end{figure*}

The above analysis shows that the frequency of coincidently fixed mutations in pairs of lineages depends on those lineages' shared clonotype. For example, the same mutation may be selected in a clonotype if the site of the mutation is involved in binding that clonotype's cognate antigen, but may be neutral in another clonotype where that site is not involved in antigen binding. For each non-synonymous mutation $m$, defined by the {V gene family} $V$ on which it occurs, its IMGT-aligned site $i$, and the amino-acid $a$ it mutates into, we can define a \textit{distribution} of fixation probabilities across different clonotypes~(\siref{si: mixture model}). We simplify this into a minimal model where each mutation $m=(V,i,a)$ can be of two types depending on the clonotype (Fig.~\ref{fig4:mixture model inference}a; Methods): beneficial for a fraction $\phi(m)$ of clonotypes, and neutral for the remaining fraction $1-\phi(m)$. When neutral, the probability of fixation is given by $p_0(m)$, reflecting SHM targeting biases {combined with drift or hitchhiking}. When selective, it is given by a constant $p_{\rm b}(m)>p_0(m)$. This constant assumption is the main simplification: in general, the selective strength of a beneficial mutation may still depend on the clonotype.  {\AW A mutation is beneficial in a clonotype if it fixes there more often than it would by drift and hitchhiking alone. }We reasoned that our data, while underpowered to measure such variation, could still capture \DW{\sout{the typical} a \textit{characteristic}} beneficial fixation probability $p_{\rm b}(m)$ \DW{for each mutation. Given there is no enrichment for coincident synonymous mutations among shared-clonotype lineages (Fig.~\ref{fig2:correlations}d), we assume that enhanced fixation probability is mostly due to the dependence on clonotype of site-specific {\em selection}, rather than of other evolutionary parameters (\siref{si: pop gen framework}).}

Within the model, we can express the fixation probability of fixation of a mutation in a lineage with a random clonotype as $\mu(m)= (1-\phi(m))p_0(m)+\phi(m)p_{\rm b}(m)$, and in two lineages with unrelated clonotypes as $\mu(m)^2$. This expression makes clear that we cannot disentangle the two contributions $\phi$ and $p_{\rm b}$ from these fixation probabilities alone. To do so, it is necessary to consider the probability of fixation of a mutation in two lineages with a shared clonotype, $w_2(m)= (1-\phi(m))p_0(m)^2+\phi(m)p_{\rm b}(m)^2$ {\TM (\siref{si: mixture model})}, which was measured in the previous sections.

\DW{Because $\mu(m)$ and $w_2(m)$ are insufficient to disentangle the three parameters $(\phi(m)$, $p_{\rm b}(m)$, $p_0(m))$,} we first estimated the neutral fixation probability $p_0(m)$ from lineages founded by out-of-frame nonproductive VDJ recombination events (Methods). Because these nonproductive lineages are linked in the cell to a productive rearrangement on the second chromosome that resulted in the functional antibody, they are driven by the same stochastic forces of genetic drift \DW{but without direct selection on mutations in the sequence itself} (SI Fig.~\ref{sfig:unproductive fixed divergence};~\citep{mccoy2015quantifying, matsen2025sitewise}). \DW{Then, for each $m$, we used the statistics of coincident fixation of $m$ among pairs of shared-clonotype to infer the two model parameters $\phi(m)$ and $p_{\rm b}(m)$ using a maximum likelihood estimator (MLE), see Fig.~\ref{fig4:mixture model inference}b and {Methods}. {The MLE is formally equivalent to matching $\mu(m)$ and $w_2(m)$ to the data} (\siref{si: mixture model}); in practice, sampling noise can cause the latter to infer unphysical parameters (e.g. $p_{\rm b} > 1$), which we avoided by constrained MLE}.
We checked that this inference was robust to our interpretation of $p_0(m)$, by showing that the inferred $\phi(m)$ and $p_{\rm b}(m)$ were similar in alternative models where non-beneficial mutations were assumed deleterious or lethal, $p_0(m)= 0$~(SI Fig.~\ref{sfig:mixture model vs p0}).

\DW{\sout{We find that most mutations $m$ have $\phi_{\rm b}$ larger than 0 but smaller than 1, meaning that they can be beneficial or neutral depending on the context. Among 234 mutations with sufficient data, 220 (94\%) have a strong  ($p_{\rm b}/p_0 > 2$) selective effect in clonotypes for which it is beneficial. The $p_{\rm b}/p_0$ ratio is typically between 10 and 100, and is larger in and around CDR1 and 2 (Fig. 4C bottom, red curve) relative to other regions (blue curve),}}
\DW{Among 299 mutations in which a beneficial class was detected after multiple hypothesis correction (FDR $q<0.05$; Methods), $243$ (81\%) have strong selective effects in clonotypes for which they are beneficial ($p_{\rm b}/p_{0}>2$). We find that most such mutations have $\phi$ larger than 0 but smaller than $1/2$, meaning that they can be beneficial or neutral depending on the context. A mutation in or around CDR1 and CDR2 frequently has a strong effect when it is beneficial, $p_{\rm b}/p_0\approx 10$--$100$, while a random FWR mutation is more likely to be broadly and weakly selected (Fig.~\ref{fig4:mixture model inference}c bottom),} even though the distribution of background neutral fixation rates $p_0$ is similar between these regions (Fig.~\ref{fig4:mixture model inference}c, top). \DW{Finite data dictates that mutations fall {\TM into an upper-right triangle of the parameter space} (white region of Fig.~\ref{fig4:mixture model inference}b, \siref{si: mixture model}). In {\TM addition, the fact that $\mu(m)$, which bounds $\phi(m) \cdot p_{\rm b}(m)$,} is always small imposes an unoccupied region at large $\phi$ and $p_{\rm b}$, {\TM suggesting} the action of long-term evolution: any mutation that was both frequently and strongly selected would have fixed in the germline long ago.} 
We estimate the total fraction of fixations {\TM of non-beneficial mutations} across all substitutions to be \DW{$\sum_m (1-\phi(m))p_0(m)/\sum_m \mu(m)\approx 55\%$}.

\DW{To validate whether the different values of $\phi(m)$ and $p_{\rm b}(m)$ across mutations reflect meaningful biological variation}, we used the model to predict the fate of mutations in held-out lineages that were not used for inference. In particular, parameter inference was restricted to pairs of lineages sharing the same clonotype, setting aside any additional lineages with the same clonotype beyond the pair. For validation, we retrieved these clonotypes that were shared by at least three distinct lineages ({in at least three subjects}), and used the model to predict the fixation probability $p_3$ of a mutation $m$ in the third lineage conditioned on its fate in the first two, using Bayes' inversion formula (see Methods and \siref{si: mixture model}).
There are three cases: $m$ fixed in none of (Fig.~\ref{fig4:mixture model inference}d), one (Fig.~\ref{fig4:mixture model inference}e) or both lineages (Fig.~\ref{fig4:mixture model inference}f). The fixation probability $p_3$ is well predicted in all three cases, and across different values of the beneficial fixation probability $p_{\rm b}$.
Next we computed the predicted fraction of clonotypes $g_{\rm ben}$ for which $m$ was beneficial, conditioned on its fate in the first two lineages and its fixation in the third. A mutation that did not fix in the first two lineages is relatively unlikely to be beneficial in that clonotype, even if it fixed in the third lineage
(Fig.~\ref{fig4:mixture model inference}d, inset).
By contrast, a mutation that fixed in at least one of the first two lineages as well as in the third was highly likely to be beneficial in that clonotype according to the model (Fig.~\ref{fig4:mixture model inference}e,f, insets). Model predictions in these conditional scenarios are largely driven by $p_{\rm b}$ (represented by the color), and independent of $\phi$ (\siref{si: mixture model}). This suggests that\DW{, aggregated across mutations, the characteristic} value of $p_{\rm b}$ is robustly inferred.  We further tested that robustness by checking that imposing a mutation-independent set of parameters, \DW{$p_{\rm b}/p_{0}=30$, $\phi = 0.02$, drawn from the modal estimate across mutations}, performed poorly (SI Fig.~\ref{sfig:pben/pneut vs pneut vs mode}a,b). \DW{While these results together establish genuine variation in the characteristic $p_b$ across mutations, it is important to note that the structure of the model inference can induce an artificial anticorrelation between $p_{\rm b}$ and $\phi$ consistent with that observed in Figs.~\ref{fig4:mixture model inference}b,~\ref{sfig:pben/pneut vs pneut vs mode}a~(\siref{si: mixture model}). An alternative test nonetheless suggests an additional, if modest, negative association (Pearson $r\,{\simeq}{-}0.35$) between the frequency and strength with which a mutation was beneficial~(Fig.~\ref{sfig:pben/pneut vs pneut vs mode}c).}

Despite its simplicity, the inference is sufficiently powered to distinguish real mutation-to-mutation variation in $(\phi, p_{\rm b})$. \DW{This variation persists within mutations of similar $p_0$, indicating that it is not only driven by SHM targeting biases (Fig.~\ref{sfig:pben/pneut vs pneut vs mode}a). The deviations in the third-lineage validation can be used to diagnose the biases in our inferences induced {\TM by our assumption of just two selection classes (neutral and beneficial)}. A more complicated structure of the true distribution of fixation rates, with the presence of deleterious mutations $p<p_{0}$ and variation in the beneficial rates, could explain some of the prediction errors~(\siref{si: model interpretation}).
  Despite these distortions, we find that these errors are frequently sufficiently small (Fig.~\ref{sfig:concealment of beneficial}) to suggest that the crude two-class model {\TM still has good predictive value. The analysis in \siref{si: model interpretation} helps to reveal when the inferred characteristic ${p}_{\rm b}$ 
  is indeed a good approximation of the typical value of fixation probability of beneficial mutations.}} Together, these results show that our coincidence-enrichment framework can meaningfully disentangle how frequently versus how strongly a particular mutation is adaptive, across the diverse clonotypes composing repertoires of healthy humans.

\subsection{Predicting statistics of convergent fixation in pathogen-specific antibodies}

\begin{figure*}[th!]
    \centering
    \includegraphics[width=1.0\linewidth]{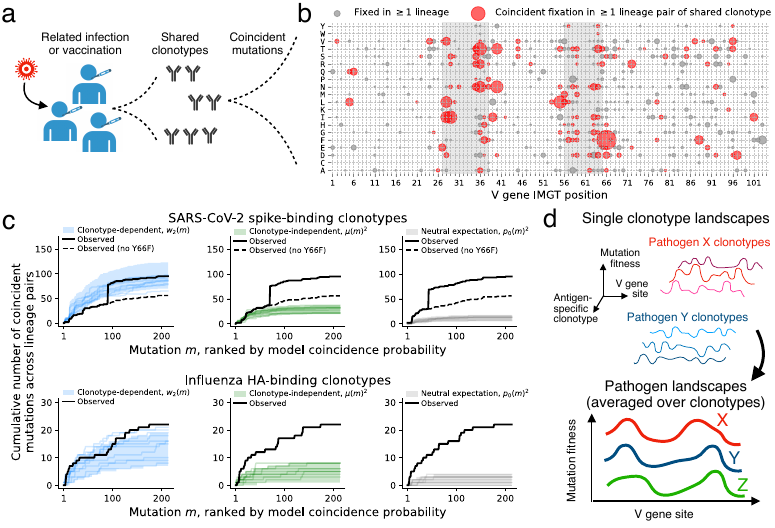}
    \caption{\textbf{Landscapes inferred from healthy individuals predict rates of convergent fixations in pathogen-specific clonotypes}. (\textbf{a}) We identify shared clonotypes among SARS-CoV-2 and influenza-specific antibodies. (\textbf{b}) Amino acid mutations 
      of SARS-CoV-2 spike protein-binding antibodies (\DW{205} public clonotypes). Grey circles indicate mutations that were observed at least once, and red circles indicate those observed in two sequences with the same clonotype. Circles are scaled by the number of lineages in which the mutation was observed, summed over all V genes.
      (\textbf{c})  Cumulative coincident fixations in shared-clonotype antibodies specific to  SARS-CoV-2 spike protein and the hemagglutinin (HA) protein of Influenza. Mutations are ranked by their inferred coincidence probability $w_2(m)$. Black thick line: data. Gray lines: simulation of the neutral model. Green lines: simulation of a naive model where each mutation is fixed with clonotype-independent probability $\mu(m)$. Blue lines: simulations of the mixture model inferred in Fig.~\ref{fig4:mixture model inference}. 10 realizations of the simulation are shown for each model. Shaded bands represent $90\%$ confidence intervals over 1000 simulations.
      (\textbf{d}) Schematic of coarse-graining from clonotype-specific to pathogen-specific adaptive landscapes. Pathogens X and Y are targeted by unique sets of clonotypes, which themselves have potentially unique sets of mutations that are beneficial. Averaging over the landscapes of pathogen-specific clonotypes defines the pathogen-specific landscape (thick lines), which themselves combine to form the global landscape~(Fig.~\ref{fig3:landscapes}).
    }
    \label{fig5:pathogen-specific antibodies}
\end{figure*}

We have so far applied our framework to affinity matured repertoires derived from healthy hosts, without prior information about the antigens targeted by these clonotypes. The parameters $\phi(m),p_{\rm b}(m)$ that we inferred characterize the statistics of beneficial mutations across clonotypes and thus across many cognate antigens. To what extent do these statistics capture the adaptive landscapes of clonotypes specific to a particular pathogen (Fig.~\ref{fig5:pathogen-specific antibodies}a)?

We retrieved around $13,100$ heavy chain sequences determined by binding or neutralization to be SARS-CoV-2 spike specific antibodies, and around $21,600$ heavy chain sequences specific to influenza hemagglutinin (HA), from multiple studies~\citep{wang2022large, guthmiller2022broadly, wang2024explainable, raju2024multiplexed, sun2026b}.
Notably, the majority of the shared clonotypes used to infer $\phi(m)$ and $p_{\rm b}(m)$ derive from subjects sampled before the SARS-CoV-2 pandemic.
We identified \DW{205} shared clonotypes (according to our earlier definition) specific to SARS-CoV-2 and \DW{99} shared clonotypes specific to Influenza HA.
From these shared clonotypes, we extracted coincident and non-coincident amino acid substitutions among the sequence pairs (Fig.~\ref{fig5:pathogen-specific antibodies}b). Some of these shared mutations, notably Y66F in SARS-CoV-2 neutralizing V3-53 antibodies~(Fig.~\ref{fig3:landscapes}c), were previously identified as recurrent and affinity-enhancing substitutions~\citep{yuan2020structural, tan2021sequence, wang2022large}.

To compare our model expectations to the observed fixation rates in these pathogen-specific datasets, we simulated the convergent evolution of all shared-clonotype sequence pairs for each pathogen.
For each shared-clonotype sequence pair, we sampled coincident fixed mutations according to three versions of the model: the mixture model from the previous section giving coincident fixation probability $w_2(m)$; a naive model where fixation probability is clonotype independent, $\mu(m)^2$; and a model where fixation probability is given by the neutral prediction, $p_0(m)^2$. These models depend on the parameters $(p_0(m),\phi(m),p_{\rm b}(m))$ inferred from healthy subjects in the previous section. Because of the small number of clonotypes,
a direct quantitative comparison of the rates of coincident fixations of individual mutations is not possible. However, counting the cumulative number of coincident fixations sorted by the model-derived rates tests if the global model carries features that remain valid for pathogen-specific mutations. 

We first sort mutations $m$ from most to least probable according to coincident fixation probability according to each model: $w_2(m)$, $\mu(m)^2$, and $p_0(m)^2$ respectively. We then plot the number of coincidently fixed mutations for SARS-CoV-2 and influenza among the $N$ most probable mutations of that sorted list as a function of $N$, for the data (Fig.~\ref{fig5:pathogen-specific antibodies}c, thick black line) as well as the results of multiple simulations of the model (thin colored lines). The neutral model (in grey) dramatically underestimates the number of fixed convergent mutations in both pathogens. The naive expectation (in green) also greatly underestimates both the SARS-CoV-2 and influenza data. On the other hand, the full clonotype-dependent model (in blue) recovers the rate of accumulation of fixed coincident mutations in both pathogens.

The global mutation landscape $(\phi(m),p_{\rm b}(m))$ used to make these predictions is a weighted average over many pathogen-specific landscapes, which themselves are averages over responses in different people, at different times and in different clonotypes (Fig.~\ref{fig5:pathogen-specific antibodies}d). This makes its predictive success on antibodies specific to a particular pathogen all the more striking. We note, however, that this cumulative prediction is not sensitive to the exact value of the inferred selection parameters: the uniform mixture model with \DW{$\phi(m)=0.02$ and $p_{\rm b}/p_0=30$} (but with $p_0(m)$ still depending on $m$) could also reproduce the data~(SI Fig.~\ref{sfig: antigen vs coarsegrain}). Further pathogen-specific data is necessary to determine to what degree the landscape is conserved across pathogens at the level of single mutations.

\subsection{Predicting clonotype-dependent beneficial mutations with antibody language models}

\begin{figure*}[th!]
    \centering
    \includegraphics[width=1.0\linewidth]{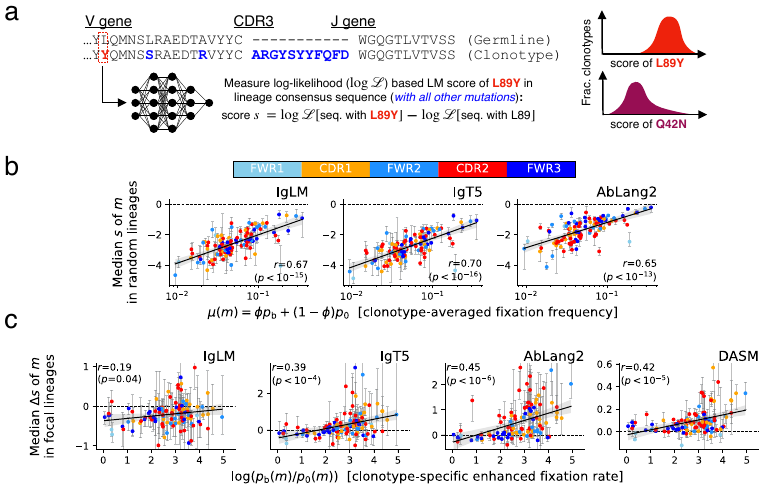}
    \caption{\textbf{Clonotype-dependent selection in antibody language models.} (\textbf{a}) Schematic: an antibody language model (LM) is used to estimate the log likelihood-based score $s$ of a mutation $m$ in a sequence as the difference of scores with and without the mutation. 
    (\textbf{b}) Median LM score, across random sequences from the repertoires, of each of 110 mutations versus its estimated fixation rate $\mu$. Error bars are $90\%$ bootstrapped confidence interval.
    (\textbf{c}) Median residual LM scores $\Delta s$ (corrected for clonotype-independent contributions, see text) versus the mutation's selective factor $p_{\rm b}/p_{0}$ inferred from our model. \DW{The same 110 mutations as in \textbf{b} are shown, representing those that were identified as conditionally adaptive ($\phi > 0$) and} coincidently fixed in at least ten shared-clonotype lineage pairs; \DW{error bars represent bootstrap confidence intervals.}
    }
    \label{fig6:language models}
\end{figure*}

Our results show that it is possible to extract signatures of clonotype-dependent beneficial selection at the level of single mutations from antibody repertoires. We next asked whether protein and antibody language models (LM), which are trained on {similar databases of B cell repertoires} using deep-learning architectures, can predict these beneficial effects. Such models have struggled with predicting particular functions~\citep{chungyoun2025fitness}, most notably binding affinity, since often antigen information is not available.
The task is made even harder when predicting experimentally generated substitutions that are unlikely to occur and thus not observed in sampled repertoires~\citep{nijkamp2023progen2}. By contrast, naturally occurring mutations in repertoires provide a more favorable benchmark to assess the predictive power of language models.

We explored several protein and antibody language models (Table~\ref{tab:language model table}), including general protein language models (ESM2~\citep{lin2023esm2}, ProGen3~\citep{bhatnagar2026scaling}) and antibody-specific language models (IgLM~\citep{shuai2023iglm},~IgBERT and IgT5~\citep{kenlay2024large}, AbLang2~\citep{olsen2022observed},  DASM~\citep{matsen2026separating}, \DW{and CoSiNE~\citep{lu2026conditionally}}).
Some recent antibody language models are ``affinity maturation (AM)-aware" (AbLang2, DASM, CoSiNE), meaning they were trained with loss objectives that suppress global biases in repertoire data, namely biases to germline residues~\citep{olsen2024addressing} and mutation biases owing to the sequence context-dependence of SHM~\citep{matsen2026separating,matsen2025sitewise}.
Almost all the antibody language models were trained on sequences derived from the Observed Antibody Space (OAS), which includes sequences derived from B19 and P22. The exception is DASM, which was trained on an independent set of repertoires from those analyzed here. We elected to keep shared clonotypes derived from OAS subjects, since they represent a large fraction of our benchmark. However, as discussed below, removing these clonotypes does not qualitatively affect our results. 

\begin{table}[h]
\centering
\begin{tabular}{|c|c|c|}
\hline
Scope & Language Model & Type \\
\hline
\multirow{2}{*}{protein LM} 
& ESM2 & MLM \\
& ProGen3 & Autoregressive \\
\hline
\multirow{2}{*}{antibody LM} 
& IgBERT & MLM \\
& IgT5 & Autoregressive \\
& IgLM & Autoregressive \\
\hline
\multirow{2}{*}{AM-aware LM} 
& AbLang2 & Germline-controlled MLM \\
& DASM & Neutral evo.-controlled \\
& CoSiNE & Neutral evo.-controlled \\
\hline
\end{tabular}
\caption{Summary and properties of the tested language models}
\label{tab:language model table}
\end{table}

We formulated the prediction task as follows~(Fig.~\ref{fig6:language models}a).
Using each language model, we evaluate the log likelihood-based score $s$ of a mutation in a particular lineage by computing the score of the lineage's consensus sequence (including all its fixed mutations), from which we subtract the score of the same sequence with the mutation of interest reverted back to germline (Methods).
This mutation score has been shown to have significant prediction capacity of a mutation's effect when applied to broad classes of non-antibody proteins
\citep{meier2021language, notin2022tranception}, and is routinely interpreted as describing the {fitness effect} of a mutation~\citep{mason2021optimization, pugh2026likelihood}. The calculation of the score depends on the nature of the language model, whether autoregressive or a masked language model (\citep{pugh2026likelihood};~Methods). The DASM \DW{and CoSiNE} models directly compute a ``selection factor" of a given amino acid in a given background sequence~\citep{matsen2026separating}, conceptually analogous to our selection factor $p_{\rm b}/p_{0}$.

Antibody language models are prone to learn averaged statistics of mutations,
namely the overall frequency of mutations across antibody sequences.
These statistics are dominated by SHM targeting biases and selection common to all clonotypes, rather than by clonotype-specific selection~\citep{olsen2024addressing, matsen2025sitewise, matsen2026separating, lu2026conditionally}.
This is confirmed within our framework: the median score $s$ over lineages of a mutation $m$ significantly correlates not only with our model's clonotype-averaged fixation rate $\mu(m)$ across antibody language models (Fig.~\ref{fig6:language models}b), but also with the neutral fixation rate $p_0(m)$ (Fig.~\ref{sfig: LM score vs pneut}a), further demonstrating that these scores capture SHM biases. By contrast, the protein language models ESM2 and ProGen3, which are not trained on antibodies, \DW{\sout{do not predict fixation at all} are not  correlated with fixation rates}. Note that antibody language models predict negative scores for most mutations, indicating a consistent preference for the germline residue that predominates the data used for training (Fig.~\ref{fig6:language models}b).

Because it is averaged over clonotypes, the overall fixation frequency of mutations does not allow us to assess the LM's ability to capture clonotype-specific selection. To do so, we must focus on lineages in which these mutations are likely to be beneficial. Concretely, for each mutation $m$,  we collected the list of shared-clonotype lineage pairs in which $m$ fixed in both lineages. Our previous analyses indicate that these coincident fixations imply that the mutation is often beneficial in that clonotype~(Fig.~\ref{fig4:mixture model inference}f). We then computed the mutation score $s$ of $m$ in each lineage from the list.
To control for the effect of the clonotype on selection, for each lineage we subtracted from $s$ the average score of the same mutation but in 20 random lineages with the same V gene and {a similar} number of fixed mutations. The resulting difference $\Delta s$ measures the clonotype-dependent contribution to the mutation score. We computed the median value of $\Delta s$ over the lineages from the list to get an overall measure of the clonotype-dependent selection strength of $m$.

This median $\Delta s$ of IgT5, AbLang2 and DASM is remarkably consistent with the logarithm of the selection factor $p_{\rm b}/p_0$ quantifying clonotype-dependent selection in our model (Fig.~\ref{fig6:language models}c), while the raw median mutation score $s$ is not (SI Fig.~\ref{sfig: LM score vs pneut}b). \DW{\sout{This was especially apparent in AbLang2 and DASM, for which}Across these models,} the median $\Delta s$ is approximately $0$ in the absence of clonotype-dependent selection, $p_{\rm b}/p_{0} \sim 1$. For IgBERT, IgT5, and AbLang2, the raw score $s$ is dominated by clonotype-independent effects such as germline and SHM bias, and was thus not expected to be a good predictor of $p_{\rm b}/p_0$. On the other hand, we expected the score $s$ of DASM to be able to predict $p_{\rm b}/p_{0}$ directly, since it was trained to remove those biases from its score. However, we found that it actually correlates negatively with it ($r{=}-0.50, p{<}10^{-7}$, SI Fig.~\ref{sfig: LM score vs pneut}b), \DW{although this is confounded by systematic inverse variation of $p_{\rm b}/p_0$ with $p_0$ (SI Fig.~\ref{sfig: LM score vs pneut}c)}. Importantly, unlike the other antibody language models, DASM was trained on independent data from the repertoires used for our inference, ruling out any data leakage between our benchmark and DASM. We also confirmed that the signal survives when restricting to clonotypes sourced from subjects not in OAS, i.e. excluding B19 and P22 subjects (SI Fig.~\ref{sfig: LM deltasrcore no OAS}a). \DW{Significantly, multiple regression of $\Delta s$ against not only $p_{\rm b}/p_0$ but also raw neutral fixation rates $p_0$ and $\phi$ shows that the correlation of $\Delta s$ with $p_{\rm b}/p_0$ is not due to confounding with these other factors (Fig.~\ref{sfig: LM deltasrcore no OAS}b), unlike the raw scores.}

We conclude that affinity maturation-aware language models are able to learn the clonotype contexts of positive selection from repertoire data alone. However, this signal can only be seen after subtracting off an appropriate null expectation that cancels out clonotype-independent biases like site-dependent SHM that dominate antibody LM learning~(Fig.~\ref{fig6:language models}c).

\section{Discussion}
Quantitatively describing the evolutionary forces operating during B cell affinity maturation is valuable to predict the ability of the adaptive immune system to respond to pathogenic challenges. A key element of these forces is the adaptive landscape that shapes the possible mutations and fixations for each individual B cell receptor. In this work, we have introduced a population genetics framework that leverages convergent affinity maturation among lineages of shared clonotype arising among different people to tease out site-level clonotype-dependent \textit{positive} selection in the V gene.

Our approach is statistical. Rather than measuring the adaptive landscape of individual antibodies, it describes a ``landscape of landscapes" across many antibodies specific to multiple antigens, and is thus complementary to existing empirical approaches to characterize single antibodies.
This ``landscape of landscapes" is broadly consistent with expectations of where strongly positively selected mutations should arise along the antibody sequence. Namely, much of the clonotype-dependent enrichment signal is clustered around the antigen-binding loops CDR1 and CDR2 (Fig.~\ref{fig3:landscapes}b). Since these regions are directly involved in antigen binding, increased fixation of specific amino acids in these regions suggests that convergent mutations are likely to affect antigen-dependent binding affinity. In contrast, strong positive selection in the framework regions is rarer, reflecting their role in stabilizing antibody structures. Beyond broad patterns, our work also uncovers more fine-scale signatures. In particular, beneficial \DW{fixation rates} are frequently strongest not within the CDR loops, but at the borders or hinges of the loops. We also uncover anomalous signatures of convergent positive selection, for instance enriched rates of cysteine fixations in some clonotypes~(SI~Fig.~\ref{sfig:more logo plots}). 
Exploring such mutations and the contexts in which they recurrently arise could provide insights into the structural and functional diversity of \DW{\sout{high-affinity} affinity matured} antibodies. Practically, our work also provides a useful prior for choosing interesting mutations in synthetic mutagenesis experiments.
For example, to produce as many positively selected mutated variants as possible one should engineer mutations with high $\phi$, whereas if one is looking for a rare but strong-effect mutation, one should prioritize mutations with high $p_{\rm b}/p_{0}$.
 
The probability of seeing the same mutation in independent lineages with the same clonotype depends on many factors: the probability that different individuals generate similar clonotypes~\citep{marcou2018high}, the probability that the same hypermutation arises and that it is selected.  As previously reported, the overall probability of these public clonotypes is well predicted by basic generation and selection models~\citep{ruiz2023modeling}. Here, we have gone beyond and shown that we can separate out the two elements that contribute to the probability of seeing a given substitution at given site in different individuals that share the same clonotype: \DW{the likelihood the mutation is adaptive, and how much this increases its fixation probability, in a particular clonotype}. This allowed us to assess whether a given mutation at a given site is actually adaptive, or just hitchhiking. Our analysis highlights that the entire clonotype sequence matters as the same mutation will fare differently on different backgrounds.

Previous work has extracted signatures of positive selection in the expanded lineages of individuals' memory repertoires by aggregating genetic signal across the antibody sequence of individual lineages~\citep{yaari2012quantifying, yaari2015mutation, horns2019signatures, nourmohammad2019fierce, hoehn2021human, ralph2020using, mikelov2022memory}. {Recent work ~\citep{mccoy2015quantifying, matsen2025sitewise, matsen2026separating} extracted clonotype-independent statistics of selection across sites by averaging genetic signals across clonotypes.} In all cases, connecting these signatures to basic evolutionary quantities, like the {\TM characteristic} strength of {clonotype-dependent} selected mutations, was out of reach. Our site-level approach, focused on public clonotypes, addresses some of these basic quantities.
For instance, \DW{\sout{we estimate} our model finds that} a {\TM characteristic} clonotype-dependent beneficial mutation arising in or near CDR1 and CDR2 has a ${\sim}10-100$-fold increased probability of fixation over a neutral mutation. Conversely, we estimate \DW{as much as $55\%$} of observed fixed amino acid substitutions \DW{\sout{are ``neutral"} fix by hitchhiking (i.e. do not fix as a beneficial mutation)}, highlighting how high rates of mutation and strong genetic hitchhiking can strongly affect the observed patterns of substitution and sequence diversity over the course of affinity maturation. 
It is tempting to relate the ratio $p_{b}/p_{0}$ to underlying population genetic parameters, like population size and the per-generation fitness effect of a beneficial mutation, that arise in minimal models of asexual evolution.
However, strong competition between different mutated clones (known as clonal interference) within B cell lineages~\citep{horns2019signatures, nourmohammad2019fierce}, and across lineages evolving in the same germinal center, complicates this mapping. 
Recent experimental work has given unprecedented quantitative insight into the evolutionary dynamics operating in single germinal centers of mice~\citep{dewitt2025replaying}. Our own work provides a timely point of comparison of the evolutionary dynamics of affinity maturation over larger length and timescales. More \DW{theoretical} work, such as fine-grained modeling of affinity maturation in individual germinal centers \DW{that incorporates clonal competition~\citep{molari2020quantitative, ralph2026inference}, as well as epistatic effects between somatic hypermutations accruing within a single lineage~\citep{pappas2014rapid, phillips2021binding, moulana2023landscape, schulz2025epistatic},} will be necessary to bridge the evolutionary dynamics from single germinal centers to the global scale studied here. 

Our analysis has relied on repertoire data from healthy individuals with no antigen information. Yet, we have shown that these statistical landscapes can reproduce rates of convergent mutations, including previously validated affinity-enhancing mutations, in public antibodies specific to both SARS-CoV-2 and influenza (Fig.~\ref{fig5:pathogen-specific antibodies}).  
However, we could not directly compare the global landscape of beneficial prevalence and selection strengths ($\phi(m),p_{\rm b}(m)$) inferred from healthy repertoires to pathogen-specific landscapes, due to limited data for the latter.
We could imagine that, owing to the precise biochemical details of binding to particular antigens, the adaptive landscape of pathogen-specific clonotypes should be idiosyncratic to that pathogen. The global statistics that we inferred should then reflect an average over many pathogens' idiosyncratic landscapes, but would not be a good approximation for any particular pathogen. Alternatively, pathogen-specific landscapes, themselves aggregating the landscapes of many antigens derived from that pathogen, could be similar to each other and thus well represented by the global landscape~(Fig.~\ref{fig5:pathogen-specific antibodies}d).
Which of these two hypotheses is correct has implications for the optimality of affinity maturation under changing viral pressures, as well as for the rational optimization of antibodies to new pathogens.
Future analysis of different antigen binding clonotypes could help us understand how the global evolutionary landscape is made up of individual antigen landscapes.
Our general inference framework could be applied to repertoires generated from individuals under acute infection or vaccination, or to repertoires sorted on specific antigen binding, to measure more directly antigen-specific adaptive landscapes.
Notably, our approach scales favorably with dataset size: a $10$-fold increase in the size of the dataset should translate into a ${\sim}100$-fold increase in the number of clonotype-paired lineages identified.

Our approach is focused on clonotype-dependent, site-level resolution of selection from repertoire data, and is thus conceptually similar to a key aim of protein and antibody language models: to evaluate mutations in their broader sequence context. Outside the realm of antibodies, protein language models have shown impressive performance at predicting mutation effect on protein function~\citep{meier2021language}. However, prediction of functional properties of antibodies has been more difficult for both protein and antibody language models. A commonly cited reason is that most antibody language models are trained on repertoire sequence data alone, which does not contain information about the antigen to which antibodies bind~\citep{chungyoun2025fitness}. However, we have shown that repertoire data alone \textit{does} contain site- and clonotype-dependent signals of beneficial mutations. \DW{We speculate that these beneficial effects are mostly attributable to antigen binding properties, although they may also involve antigen-independent phenotypes like protein stability via interactions between the CDR3 loop and residues of the V gene}. We were able to detect the same signal in antibody language models, especially those that correct for germline and SHM biases~\citep{olsen2024addressing, matsen2025sitewise, matsen2026separating}, but only after isolating clonotype-independent effects through $\Delta s$. {Our work reinforces the value of strong biological priors to guide data curation, training, and testing in modern machine learning methods~\citep{matsen2025sitewise, matsen2026separating}.}

A principal shortcoming of our approach is that it relies on data that provides only a very partial view of each observed clonotype's adaptive landscape, namely the mutations that arose and fixed in pairs of lineages. Nonetheless, since SHM biases drive more mutations to arise in the binding-related CDR1 and CDR2 loops, we expect our framework to provide good coverage of mutations that tend to affect antigen binding. Thus, despite its narrow view into any particular clonotype's landscape, our framework still captures a significant fraction of the typical adaptive mutations arising in the V gene.

\section{Methods}
\subsection{Data processing and intra-subject lineage construction}
Antibody repertoire datasets were downloaded from the following locations:
B19~\citep{briney2019commonality} at \url{https://github.com/brineylab/grp-paper};
P22~\citep{phad2022clonal} at European Nucleotide Archive (ENA), accessions E-MTAB-11174 and E-MTAB-11697;
M22~\citep{mikelov2022memory} at ENA accession E-MTAB-11193; and C25~\citep{cvijovic2025long} at NCBI BioProject database accession PRJNA1203354. Only the 2 most deeply sequenced individuals (TBd5, TBd6) out of 6 from C25 were included. Details about the size of each datasets are summarized in SI Table~\ref{tab:subject_stats} (SI \siref{si:datasets}).

Heavy chain sequences were processed following the pipeline of the original studies, and through \texttt{igblastn} v1.22.0~\citep{ye2013igblast} with the IMGT germline library for alignment and VJ annotation. For bulk data, UMI-dereplicated sequences represented by less than 3 reads were excluded to reduce errors. Identical sequences with the same isotype were merged in each bulk sample to avoid double counting cells. Samples from the same individual were then merged into a single dataset.
To account for polymorphism in the germline V genes, we used TIgGER~\citep{gadala2015automated} to construct subject-specific germline V alleles, when possible, using a maximum of a million sequences per subject. For M22 and P22, we only kept memory and plasma cells. For all datasets, for consistency we only analyzed memory sequences defined as having at least 3 mutations. We used HILARy~\citep{spisak2024combining} to cluster productive (no frameshift in CDR3) sequences into putative lineages. The lineage's clonotype is defined by its V and J genes as well as the consensus CDR3 amino acid sequence. We also clustered nonproductive memory sequences into nonproductive sequences for B19 and M22, which contained enough such sequences. More details about the data processing procedures can be found in \siref{si:sequence processing}.

\DW{As noted in the main text, a single nucleotide mutation was considered fixed in a lineage if it was observed in ${\geq}80\%$ of de-replicated sequences in the lineage. Fixed amino acid mutations were similarly defined, without requiring that the underlying codons be identical (to allow additional synonymous mutations to accumulate in that codon). Details about the macroevolutionary statistics and supplementary analyses involving these and related quantities (Figs.~\ref{sfig:macroevo correlations}, \ref{sfig:gene specific macroevo correlations}) are reported in SI \siref{si: evolutionary summary statistics}.}

\subsection{Defining inter-subject clonotype sharing}
\label{si:clonotype sharing}
Lineages of the same VJ$\ell$ class were aggregated across subjects. Metaclusters of shared clonotypes were constructed by single-linkage clustering at $85\%$ amino acid identity among lineage consensus CDR3s. Single-linkage clustering produces a deterministic set of metaclusters which were convenient for bookkeeping. However, this procedure also groups lineages with consensus CDR3 identities ${<}85\%$. For all downstream analyses, subsets (e.g. pairs) of lineages were subsampled from a metacluster as a shared clonotype only if all pairs of lineages in the subset satisfied this $85\%$ identity threshold. Metaclusters also frequently contained multiple lineages from the same subject. To mitigate the effects of spurious mutation correlations if the within-subject lineage clustering step incorrectly identified a single lineage as two independent lineages (perhaps divided by a deep branch), no more than one lineage from the same subject was ever included in a subset of a shared clonotype. 

\subsection{Prediction of \DW{\sout{the amount of clonotype sharing}} clonotype sharing with synthetic repertoires}
We used OLGA~\citep{sethna2019olga} and SoNNia~\citep{isacchini2021deep} to generate synthetic clonotypes to predict the amount of expected sharing by chance. OLGA models the raw recombination process, and was used with default (pre-trained) parameters. SoNNia is built on top of OLGA and models initial and global peripheral selection on antibodies. We trained a selection model on each subject dataset by first sampling up to 1 million sequences from the germline-aligned and lineage-clustered libraries. Among B19 subjects, sampling was further restricted to only two or fewer mutations in the templated (non-CDR3) regions. These lightly mutated sequences were not represented in cross-subject sharing analyses, and thus represented an adversarial training set to compare against observations. (The same was not possible for the other datasets, which were biased toward memory or plasma cells with high mutation counts, Fig.~\ref{sfig:subject repertoire statistics}f.) A subject-specific selection models  was then trained on sampled sequence data, using the mutiple-layer feature of SoNNia only when the number of sequences was at least $10^5$ (as per~\citep{isacchini2021deep}).

For each subject, a synthetic selected repertoire was generated, of size equal to the number of filtered lineages as in the real data. Each lineage was then randomly associated with a lineage size drawn from the same subject's lineage size distribution. Clonotype sharing was then estimated exactly as for the real data.

\subsection{Computing coincident fixations of mutations in paired lineages}
\label{si: landscape of coincident mutations}
\newcommand{\ellkone}{\ell_{k,1}}
\newcommand{\ellktwo}{\ell_{k,2}}

\DW{
{\TM We estimate $w_2(m)$ as the fraction of paired lineages for which $m$ is fixed in both lineages, among the lineage pairs that share their clonotype and in which $m$ could have arisen.} Fig.~\ref{fig3:landscapes}a plots $\hat{w}_2 (m)$, summed across $m$ of different $a$ but sharing the same family $V$ and IMGT position $i$.} Figures~\ref{fig2:correlations}c,d plot statistics of total counts of coincident fixation to the same $a$ at the same $i$, aggregated across IMGT sites in each pair \DW{(excluding the CDR3 and primer-masked regions)}. In all cases, only paired lineages with at least one fixed amino acid change in each lineage were retained for calculations.

The null coincidence rates (Fig.~\ref{fig2:correlations}c,d and Fig.~\ref{fig3:landscapes}a) are calculated as above, but with a set of clonotype-unmatched lineage pairs.  \DW{\sout{This set $S^{(null)}$ of unmatched lineage pairs were constructed beginning with the original observed set of lineage pairs as follows. For the $k$th pair in $S$, a new pair was created by preserving the first lineage $\ellkone$. The second lineage $\ellktwo^{(null)}$ was randomly drawn among lineages from}}
\DW{This set was constructed beginning with the original observed set of lineage pairs as follows. For the $k$th pair among $N_m$ ordered lineage pairs of shared clonotype {\TM in which $m$ could have occurred}, a new pair was created by preserving the first lineage $\ellkone$ while substituting the second $\ellktwo$. The second was replaced with a ``null'' lineage $\ellktwo^{(\rm null)}$, randomly drawn from lineages (i) belonging to the same subject and V gene as $\ellktwo$; (ii) having a measured abundance within an order of magnitude of $\ellktwo$, $0.1\leq|\ellktwo^{(\rm null)}|/|\ellktwo| \leq 10$; and (iii) with a similar amino-acid divergence from the germline V region $D_{\rm aa}$, {\TM $D_{\rm aa}(\ellktwo)<D_{\rm aa}(\ellktwo^{(null)}) < \max [ 1.5\cdot D_{\rm aa}(\ellktwo), D_{\rm aa}(\ellktwo)+2]$.}} This procedure preserves gene-specific mutation rates and local sequence context, and conserves (indeed over-estimates) the joint mutation load within each pair, and thus the per-pair probability of coincident mutation. \DW{\sout{To the extent that measured variation in lineage sizes reflects difference in evolutionary dynamics and thus on mutation fixation probabilities (see below), the constraint on size seeks to preserve these statistics between the observed and null pairs.}Restricting to similar abundances aims to control for any variation in evolutionary dynamics and thus on mutation fixation probabilities (\siref{si: pop gen framework}).} Defining the mutation-specific null coincidence rate $w_2^{\rm (null)}(m)$, Fig.~\ref{fig3:landscapes}b plots the enrichment ratio $\hat w_2(m)/\hat w_2^{\rm (null)}(m)$.

\subsection{Model inference with lineage pairs of shared clonotype}

{\TM For each mutation $m$, once $p_0(m)$ is fixed using nonproductive lineages, our two-class model has 2 free parameters, $\phi(m)$ and $p_{\rm b}(m)$, which we infer using a maximum likelihood estimator (MLE) using pairs of clonotype-sharing lineages in which $m$ is accessible (i.e. have the same $V$ as $m$).
  For each clonotype $\sigma$ shared by a pair of lineages, and mutation $m$, call $k_\sigma(m)=0,1,2$ the number of lineages in the pair in which $m$ is fixed. The likelihood of the data given the parameters is:
  \begin{equation} \label{eq: methods log-likelihood}    
  \begin{split}
    \mathcal{L} 
    &= \sum_{\sigma} \ln \bigg[  \phi(m) p_{\rm b}(m)^{k_\sigma(m)} (1- p_{\rm b}(m))^{2-k_\sigma(m)}  \\
    &+ (1-\phi(m)) p_{0}(m)^{k_\sigma(m)} (1- p_{0}(m))^{2-k_\sigma(m)} \bigg] + \textrm{const.}
    \end{split}
\end{equation}}
{\TM To avoid unphysical estimates of either $\hat{\phi}$ and $\hat{p}_{\rm b}$ outside the domain $[0,1]$ owing to finite sampling, we used the constrained MLE estimates on the domain $(\phi, p_{\rm b}) \in [0,1]^2$.}

{\TM A mutation $m=(V,i,a)$ was considered for mixture model inference if (i) it had ${\geq}100$ pairs of clonotype-sharing lineages with the right $V$ (additional lineages from triplets and more were left out and used in the cross-validation described in the next section); and (ii) the estimated $p_0$ and ${\mu}$---both independent of lineage pair data---were sufficiently well-measured {\TM(less than 50\% relative standard error)}. The mixture model was fit for 1259 mutations that met both criteria. The model fit was considered to have identified a beneficial class if \DW{\sout{the estimated excess ${\mu}-p_0$ was significantly above 0} (i) the excess ${\mu}-p_0$ was greater than zero and (ii) the likelihood ratio of the two-class fitted model to the neutral model was large relative to the likelihood ratios of $20,000$ parametric bootstrap resamples of pair coincidence counts under neutrality} (299 mutations; Benjamini-Hochberg adjusted $q<0.05$ across 1259 mutations).
Of those, 86 mutations were consistent with $\phi=1$ and $p_{\rm b} = \mu$ \DW{under a similar likelihood ratio test as above, but with 1-class parametric bootstrap resamples centered on $p = \mu$}.
127 mutations with positive but insignificant excess $\mu-p_0$ were then assumed to be always neutral and set to $\phi(m) = 0$. The rest of mutations, with $\mu<p_0$, correspond to net purifying selection that is inconsistent with our model and makes ruling out a positive class difficult to do from pairwise coincidence data alone. Where plotted, errors represent the interquartile range of 1000 bootstrap resamples. \DW{The lower sampling-constrained bound is defined using the sampling depth $N(m)$ and externally fixed $p_0$ across all 1259 mutations, and its exact form is derived in~\siref{si: mixture model}. Practically, the bound amounts to $p_{\rm b}(\phi)=\sqrt{\frac{2}{\max_m N(m) \phi}}$.}
}

\subsection{Predicting fixation in third held-out lineages}
Under the mixed model, the probability that a mutation $m$ fixes conditioned on its fixation status in two other lineages with the same clonotype $\sigma$ reads:
\begin{equation}
  p_{3}=\frac{ \phi p_{\rm b}^{1+k}(1-p_{\rm b})^{2- k} + (1-\phi)p_0^{1+ k}(1-p_0)^{2- k}}{\phi p_{\rm b}^{ k}(1-p_{\rm b})^{2- k} +(1-\phi)p_0^{ k}(1-p_0)^{2- k} },
\end{equation}
where $k=0,1,2$ is the number of lineages among the first two where $m$ was fixed, and where we have dropped $m$ dependencies of $p_3(m|k)$, $\phi(m)$, $p_{\rm b}(m)$ and $p_0(m)$ for brevity. The contribution of beneficial mutations to that probability reads:
\begin{equation}
g_{\rm ben}=\frac{ \phi p_{\rm b}^{1+k}(1-p_{\rm b})^{2- k}}{ \phi p_{\rm b}^{1+k}(1-p_{\rm b})^{2- k} + (1-\phi )p_0^{1+ k}(1-p_0)^{2- k}}\,.
\end{equation}
More detailed analysis may be found in \siref{si: mixture model}.
\subsection{Comparison to antigen-specific antibodies}
\label{si:antigen-specific antibodies}
Heavy chain sequences of SARS-CoV-2-binding clonotypes were taken from a meta-study by Wang~\etal~\citep{wang2022large}, which collected data from multiple studies which isolated monoclonal antibodies either binding or neutralizing SARS-CoV-2 (via the spike protein in ${>}99\%$ of cases~\citep{wang2022large}). This dataset was augmented with CoV-AbDab~\citep{raybould2021cov} sequences deposited after the Wang metastudy. Influenza HA-binding heavy chain sequences were taken from a similar meta-study by the same authors~\citep{wang2024explainable}. This data was augmented with data from three additional studies in which single-cell sequencing libraries of HA-binding cells were collected after recent flu vaccination and/or infection~\citep{guthmiller2022broadly,raju2024multiplexed,sun2026b}.

For each pathogen-specific set of antibody heavy chain sequences, we used IgBlast~\citep{ye2013igblast} to align all sequences to the IMGT germline database. For the subset of clones that had no accompanying nucleotide sequences, we elected to directly align protein sequences. While some sequences derived from single donors, most sequences derived from libraries representing multiple donors. Treating these libraries as ``pseudo-donors", we identified (pseudo-)lineages, represented by one or more sequences sharing the same (IMGT) annotated V gene, J gene, and CDR3 hamming distance ${<}15\%$ in each donor or pseudo-donor (with the expectation that sequences belonging to the same pseudo-lineage derive more frequently than not from the same individual within the pool of donors). To maximize the number of shared clonotypes identified, lineages of single sequences were also retained. Affinity-immature lineages lacking any amino acid substitutions in their sequences were discarded. We used the same V gene, J gene, and CDR3 Hamming distance threshold to identify shared clonotypes represented by demonstrably independent lineages deriving from different hosts. When greater than two lineages represented a clonotype, lineages were then split into randomly assigned pairs. 

Among the resulting pairs of \DWstrike{expanded} lineages of shared clonotype, we counted the number of fixed mutations and the number of coincident fixed mutations (to the same amino acid, at the same site) among pairs~(Fig.~\ref{fig5:pathogen-specific antibodies}b). We compared these results to the following simulation-based expectation parameterized by our model inferred from healthy repertoires. For each clonotype represented among the pathogen-specific shared clonotypes, we randomly assigned the effect of a given mutation $m$ as beneficial in the clonotype according to the associated model probability $\phi(m)$, provided that this mutation was \emph{accessible} to the clonotype, meaning that it belonged to the mutation's $V$ family and the substituted amino acid was within one nucleotide change of the germline allele at that site. For simplicity, we assumed that association in the beneficial class ($\phi(m)$) for $m$ at one site was uncorrelated with association in the corresponding class $\phi(m')$ for another mutation $m'$ at the same site. For each pair of lineages with that clonotype, we then sampled a coincident fixation of $m$ with probability $p_{2}(m)=p_{\rm b}^2$, if the clonotype was assigned to the beneficial class for that mutation, or $p_{2}(m)=p_0^2$ otherwise. When more than one mutation at site $i$ was accessible to the clonotype $\sigma$, the probability of coincident fixation of one of these mutations was taken as the sum over their individual (clonotype-dependent) coincident fixations, $P_2(i) = p_{2}(m) + p_{2}(m') + \ldots$ (with a corresponding probability of no coincident fixation $=1-P_2(i)$). The specifically assigned mutation $m$ was sampled according to its weight $p_{2}(m) / P_2(i)$. This procedure was carried out for all accessible mutations for all clonotypes represented by each pathogen, and Fig.~\ref{fig5:pathogen-specific antibodies}c reports the outcomes of 1000 runs per pathogen.
 
\subsection{Antibody language models}
\label{si:language models}

The suite of protein and antibody language models studied here span different training data,  architectures, and objective functions~(Table~\ref{tab:language model table}). The majority of antibody language models are trained on the Observed Antibody Space~\citep{olsen2022observed}, which includes B19 and P22 data. The exception is the DASM model, which was trained on independent repertoire data.

We followed the standard definition of mutation ``fitness" scores for both autoregressive and masked language models, respectively~\citep{pugh2026likelihood}. For any sequence $x$, denote a reference sequence $\tilde {x}$ as the sequence without the mutation $m$. For autoregressive models, the fitness score of the mutation was defined as 
\begin{align}
\label{seq: autoregressive fitness}
    s(m|{x}) = \sum_j \big [ \log p(x_j |x_{<j}) - \log p(\tilde x_j|\tilde x_{<j}) \big ].
\end{align}
For masked language models, the score was defined as
\begin{align}
\label{seq: mlm fitness}
  s(m| {x}) = \log p(x_i |x_{\backslash i}) - \log p(\tilde x_i|x_{\backslash i}),
\end{align}
where $i$ is the site of $m$. \DW{The ProGen3 score uses the average of forward (N$\to$C) and reverse (C$\to$N) autoregressive passes (Eq.~\ref{seq: autoregressive fitness}), following the convention for this model.}
Finally, the DASM and CoSiNE models directly compute a ``selection factor" for every mutation $m$, given the reference amino acid sequence $\tilde x$. \DW{The CoSiNE model additionally requires the nucleotide reference sequence to compute the SHM null.

Multiple linear regression of LM scores against the features $\log p_0$, $\log \hat{\phi}$, and $\log \hat{p}_{\rm b}/p_0$ (and allowing for a non-zero intercept) was performed using all adaptive-identified mutations with at least five lineage pairs of shared clonotype in which coincident fixation was observed. }

{\TM
  \subsection{Data availability}
  Data used in this paper can be found at: \url{https://github.com/brineylab/grp-paper}; European Nucleotide Archive (ENA), accessions E-MTAB-11174, E-MTAB-11697 and E-MTAB-11193; NCBI BioProject database accession PRJNA1203354.
The code supporting this manuscript, from pre-processing through figure generation, is available at \url{https://github.com/statbiophys/survey-10000-antibodies} upon manuscript publication.
}

\section*{Acknowledgements}
This work was supported by the ERC Synergy Grant no 101224126, ERC PoC grant no.
927 101185627, the Foundation pour la Recherche Medicale grant Team
Project EQU202503019997, the IA CLUSTER 2023 PROJET PR[AI]RIE-PSAI grant, the ANR-24-CE45-7957 grant,  the EvoMG
HORIZON-MSCA-2024-DN-01-01 and the CZI Theory Initiative grant.
{\TM We thank the anonymous reviewer who suggested a closed-form solution to our inference problem, which led us to greatly improve the interpretation of our results.}

\bibliographystyle{pnas.bst}

\appendix

\clearpage

\onecolumngrid

\setcounter{figure}{0}
\setcounter{table}{0}
\renewcommand{\thefigure}{S\arabic{figure}}
\renewcommand{\thetable}{S\arabic{table}}
\newcommand{\sub}{\mathbf{s}}

\section{Deep repertoire sequencing datasets}
\label{si:datasets}
 
The B cell repertoire datasets analyzed in this study fulfilled three criteria. Each study (i) amplified and sequenced the (nearly) complete length of the variable region; (ii) incorporated unique molecular identifiers (UMIs) to mitigate false identification of sequence diversity; and (iii) achieved relatively high per-subject sampling depth, given its respective (bulk or single cell) sequencing modality. The table below summarizes the size of the datasets for each subject.

\begin{table}[h]
\centering
\begin{tabular}{llrrr}
\hline
Dataset~~ & Subject~~ & \makecell{Filtered~~~\\Seq. Depth~~} & \# Lineages~~ & \makecell{\# Lineages\\(size${\geq}$2)}\\
\hline
briney19 & 316188 & 849,616 & 374,693 & 87,526 \\
 & 326650 & 2,912,968 & 2,238,941 & 138,197 \\
 & 326651 & 8,769,067 & 5,014,804 & 438,496 \\
 & 326713 & 8,432,057 & 5,442,852 & 457,715 \\
 & 326737 & 1,143,280 & 798,379 & 111,589 \\
 & 326780 & 3,614,612 & 1,769,724 & 334,132 \\
 & 326797 & 2,217,862 & 1,396,900 & 140,950 \\
 & 326907 & 1,067,463 & 740,885 & 75,777 \\
 & 327059 & 4,004,509 & 2,353,457 & 221,284 \\
 & D103 & 631,744 & 473,153 & 57,507 \\
\hline
phad22 & Donor1 & 246,433 & 52,545 & 31,958 \\
 & Donor2 & 460,641 & 69,322 & 64,919 \\
\hline
mikelov22 & AT & 94,887 & 54,963 & 12,824 \\
 & D01 & 149,114 & 85,869 & 20,161 \\
 & IM & 136,661 & 81,388 & 17,773 \\
 & IZ & 170,009 & 105,461 & 21,620 \\
 & MRK & 106,215 & 67,389 & 12,747 \\
 & MT & 38,610 & 26,331 & 4,952 \\
\hline
cvijovic25 & TBd5 & 121,057 & 79,070 & 16,916 \\
 & TBd6 & 83,488 & 51,122 & 10,921 \\
\hline
\end{tabular}
\caption{Subject summary statistics.}
\label{tab:subject_stats}
\end{table}

\section{Sequence alignment, annotation, and memory lineage clustering}
\label{si:sequence processing}

 \textbf{Initial alignment.} B19 furnished aligned and UMI-dereplicated sequences. For the bulk VDJ sequencing libraries of M22, we used the end-to-end \texttt{analyse} function of the MiXCR~\citep{bolotin2015mixcr}, configured with the same parameters as those of the original authors of the data, available as the preset ``mikelov-et-al-2021" in MiXCR (\url{mixcr.com/mixcr/reference/overview-built-in-presets}). Among these two bulk studies, we excluded UMI-based consensus sequences supported by ${<}$3 reads in all downstream analyses to filter out PCR and sequencing errors. For both P22 and C25, we used 10X Genomics CellRanger v9.0.1 with default settings (both original studies also used this pipeline). All productive, full-length heavy chain sequences identified by CellRanger were retained, without requiring a paired light chain, consistent with the structure of the unpaired bulk data. To standardize alignments and VDJ annotations, every library was ran through \texttt{igblastn} v1.22.0~\citep{ye2013igblast} with the IMGT germline library. \\

 \textbf{Merging libraries across samples.} Every study sequenced each subject across multiple biological samples. In B19, samples were ``biological replicates" from blood, and were sequenced as unsorted mixtures of naive and memory cells~\citep{briney2019commonality}. In C25, unsorted samples were sourced from multiple organs, including memory-enriched spleen and lymph node~\citep{cvijovic2025long}. In M22 and P22 sampling spanned over 1 and 10 years, respectively, and memory cell-specific libraries were obtained by cell sorting~\citep{mikelov2022memory, phad2022clonal}. These differences in sampling notwithstanding, all libraries in a subject were naively concatenated to build the largest memory repertoire in each subject. In the bulk sequencing context (B19 and M22), identical sequences could derive from the same cell (particularly in the case of RNA sequencing, where each cell may be represented by $\gg 1$ template). Thus, to prevent false identification of ``expanded" lineages in bulk-sequenced subjects, consensus sequences in each library that were only distinguished by their associated UMIs, but shared identical VDJ sequence and isotype, were collapsed. By the same logic, identical sequences deriving from technical replicates from the same biological sample (same extracted RNA) were also collapsed. Among single-cell sequenced subjects (P22 and C25), identical VDJ sequences within or between libraries were not merged, since these sequences necessarily derive from different cells and thus encode information about lineage abundance. \\

 \textbf{Realignment to subject-specific germlines.} Subject-specific germline V alleles were constructed using TIgGER~\citep{gadala2015automated} from unique sequences. Denoting the total number of sequences in subject $S$ by $D_S$, $\min\{10^6,D_S\}$ sequences were used for germline inference. For all V genes in the IMGT database for which TIgGER could not identify subject-specific alleles, the subject's database was supplemented with all IMGT alleles representing those V genes. Sequences were then re-aligned to their subject-specific germlines using IgBLAST. \\

 \textbf{Filtering putative memory cell sequences in unsorted repertoires.} In unsorted datasets B19 and C25, concatenated libraries were filtered down to sequences with $\geq 3$ mutations from the germline V gene, as a mutational proxy for belonging to the memory subrepertoire. While exclusively affinity matured memory and plasma cells were sampled in M22 and P22, for consistency with the unsorted libraries, sequences in both single cell datasets were also included for analysis only if they harbored $\geq 3$ mutations from the germline V gene. \\

 \textbf{Constructing memory lineages.} For each subject, sequences were sorted into ``VJ$\ell$" groups sharing the same V gene and J gene assignments and same CDR3 length. Within each VJ$\ell$ group, we used HILARy~\citep{spisak2024combining} to single-linkage cluster sequences into putative lineages, under the following parameters: \texttt{full-method --precision 0.99 --sensitivity 1}. 
In the ultradeep sequencing contexts considered here, clustering could in principle be composed of cells with two or more naive ancestors, established by recurrent recombination, though in general one might expect that these over-clustered cases should be dominated by cells from only one of the lineages. With this caveat, we proceed with describing each within-subject cluster as a single lineage. 
Each lineage was labeled with a consensus CDR3 defined as the sequence constructed from the most common nucleotide at each site, across all sequences in the lineage. \\

 \textbf{Constructing neutrally evolving lineages from nonproductive out-of-frame sequences.} Putative neutrally evolving lineages were constructed from sequences representing nonproductive out-of-frame VDJ recombination events. In particular, nonproductive sequences whose CDR3 was out-of-frame were aligned to the subject-specific germline alleles established with productive sequences. Filtering for putative ``memory" sequences with ${\geq}3$ mutations in the templated regions, lineages were then constructed using the approach described above. Only B19 and M22 carried sufficient out-of-frame sequences for lineage reconstruction. 

\section{Defining fixed mutations and other evolutionary summary statistics}
\label{si: evolutionary summary statistics}
\DW{\sout{The $80\%$ threshold frequency used in this study to annotate a fixed mutation is less than the nominal 100\% definition of fixation. Our choice allows for high-frequency mutations to be included, since, from a population genetic perspective, these mutations behave similarly to genuinely fixed mutations (see below). It also ensures genuinely fixed mutations will still be detected despite potential PCR amplification errors or misclustering of lineages. Practically, for the vast majority of lineages, represented by $\leq 4$ sequences in the data, the $80\%$ threshold is indistinguishable from fixation (${=}100\%$). }}

For each lineage $\ell$, we defined the following macroevolutionary statistics:
\begin{itemize} 
    \item \textbf{Lineage abundance $|\ell|$:} the number of (deduplicated) sequences.
    \item \textbf{Nucleotide divergence $\overline{d}_{\rm nt}(\ell)$:} the average number of nucleotide mutations from germline in the templated (non-CDR3) regions. The CDR3 is excluded since, in general, it is difficult to confidently assign germline residues to most of the CDR3, particularly in lineages represented by few sequences.
    \item \textbf{Synonymous divergence $\overline{d}_{\rm syn}(\ell)$:} analogous to $\overline{d}_{\rm nt}(\ell)$, restricted to synonymous nucleotide mutations. 
    \item \textbf{Amino acid divergence $\overline{d}_{\rm aa}(\ell)$:} analogous to $\overline{d}_{\rm nt}(\ell)$, considering amino acid mutations.
    \item \textbf{Fixed nucleotide $D_{\rm nt}(\ell)$:} the number of fixed ($\geq80\%$) nucleotide mutation.
    \item \textbf{Fixed synonymous divergence $D_{\rm syn}(\ell)$}: Analogous to $D_{\rm nt}(\ell)$, restricted to synonymous mutations.
    \item \textbf{Fixed amino acid divergence $D_{\rm aa}(\ell)$}: Analogous to $D_{\rm nt}(\ell)$, considering amino acid mutations.
\end{itemize}

 The observed correlations per subject pair were compared to a permutation null. This null was constructed by permuting the labels of lineages belonging to the same VJ$\ell$ class within each subject, and then recomputing the macroevolutionary correlations. For each matrix of pairwise correlations in~SI Figs.~\ref{sfig:abundance correlations},~\ref{sfig:macroevo correlations},~\ref{sfig:gene specific macroevo correlations}, a single label permutation was performed for each subject. This preserves cross-correlations between different pairs (sharing one subject) in the null, as in the observed data.

\section{Population genetic framework of coincident fixations}
\label{si: pop gen framework}
This section motivates the coincidence statistics in a theoretical framework that is sufficiently general to account for the (unknown and/or unmeasured) complexities of the evolutionary histories represented by memory B cell lineages. 

Concretely, we are interested in the probability that a particular mutation $m$ in a lineage of clonotype $\sigma$ arises and fixes by the time of observation. This probability depends, first, on the rate of mutation at that site, which is highly dependent on local sequence context~\citep{yaari2013models, elhanati2015inferring, spisak2020learning}. Next, fixation depends on the adaptive effect of the mutation itself, which may be a function not only of the clonotype of the lineage, but also of other mutations acquired by the lineage during the course of affinity maturation. Finally, fixation will also be a function of the larger population and evolutionary dynamics of the lineage. For instance, random demographic fluctuations within the lineage, or ``genetic drift," can drive even a neutral mutation to high frequency and fixation. A mutation may also ``hitchhike" to fixation if it is linked in the same genome with another adaptive mutation(s). The strength of these random forces depend on the adaptive landscape within a clonotype, as well as the ecological dynamics of the focal lineage as it competes with other lineages in the germinal center. 

\subsection{Model}
We assume the following minimal population genetic model of lineage dynamics. Each lineage is represented by a single founder cell that entered a germinal center (GC) reaction at time $t=0$. The total population size of the lineage within the GC is given by $N(t)$, which, as noted above, depends on the mutations acquired by the lineage and competition with other lineages in the GC. The GC reaction continues for a random time $t_{GC}$, typically a few weeks in humans, during which the \DW{antibody sequence (but not other parts of the B cell's genome)} mutates by SHM. After $t_{GC}$, the GC dissipates, SHM is turned off, and the surviving (affinity matured) cells representing the lineage exit and continue to be governed by stochastic lineage dynamics without mutation~\citep{desponds2016fluctuating}, until a sampling time $t$. This simplified picture, which is often implicitly applied in B cell phylogenetics~\citep{hoehn2021human}, allows us to apply classic population genetic results derived for well-mixed populations.

The above neglects exit of lineage cells from the GC before $t_{GC}$; this will tend to reduce observed fixation rates, since these cells cannot further mutate. One could formally model this by leakage into a non-mutating class. The resulting fixation probability will have a complicated dependence on the population size of the lineage over time in the GC, and on the difference in long-term (post-GC) survival of memory lineages leaving relatively early (less mutated) and leaving relatively late (more mutated and/or differentially ``fit"). Alternatively, one can assume that there is an intermediate time during the GC reaction, when the lineage is contributing the highest number to leaked memory cells: this will occur late enough for the lineage to have reached a large size (exponentially larger than its founding size), but not too late to begin to typically exit as short-lived plasma cells~\citep{cvijovic2025long}. Provided this intermediate time contribution dominates the post-GC population, the leakage model is equivalent to the simplified model above, with a $t_{GC}'< t_{GC}$. If this regime is not satisfied by a typical memory lineage, then the exact forms of the results below will not necessarily apply. However, the qualitative conclusions of the model, in particular how to interpret neutral versus non-neutral fixation and clonotype-conditioned enrichment in fixation, will not depend on these model details.

\subsection{Neutral mutations}
 Provided $m$ is a synonymous neutral mutation whose mutation does not itself affect the rate of (adaptive) mutations at nearby sites (although this is probably sometimes violated given the strong sensitivity to local sequence context of SHM), the neutral mutation cannot directly affect the evolutionary dynamics of a clonotype. This non-dependence greatly simplifies the analysis, in particular eliminating dependence on the population dynamics $N(t)$. Classic results from population genetics show that, neglecting back mutations, the probability $p_{\rm fix}$ that a neutral mutation \textit{destined to fix} arises is
\begin{align}
\label{seq: minimal model}
    p_{\rm fix,neutral}(m|\sigma) = \langle 1 - e^{-r t_{\rm GC}}\rangle_{t_{\rm GC}|\sigma},
\end{align}
where $r$ is the rate ({per cell per unit time}) with which $m$ arises. This is not strictly the same as the probability that the mutation has fixed in the population by time $t_{\rm GC}$, since it takes a finite amount of time to rise from one individual in the lineage to fixation.
However, in the present model, we observe the lineage at a time $t \gg t_{\rm GC}$. Provided the post-GC time $t - t_{\rm GC}$ is sufficiently large relative to the post-GC fixation time, then Eq.~\ref{seq: minimal model} approximately equals the probability of the mutation reaching fixation by that time. Alternatively, if one assumes that lineage dynamics are slow / frozen after the GC, then any mutation must (nearly) fix while the lineage remains in the GC. The same form applies with a modulated $t_{\rm GC}' < t_{\rm GC}$, roughly less than the coalescent/fixation time within the GC.

Considering the rates of coincidence, two lineages with the same clonotype may have different $t_{GC}$, owing to intrinsic variation in the lifetime of GCs or in the entry time of the founder cell. Two lineages with different clonotype may have $t_{GC}$ that vary \textit{in distribution}, for example if they respond to pathogens with different GC kinetics. This, in principle, could be reflected in heightened fixation correlations among pairs of lineages of shared clonotype:
\begin{align}
    w_{2,\rm neutral}(m) = \langle p_{\rm fix,neutral}(m|\sigma)^2\rangle_\sigma \geq \langle p_{\rm fix,neutral}(m|\sigma)\rangle_\sigma^2=p_0(m)^2. 
\end{align}
However, the agreement of coincident synonymous fixation rates among lineages with or without shared clonotypes (Fig.~\ref{fig2:correlations}d, Fig.~\ref{sfig:vgene panel different datasets}) imply that this is a weak effect, $w_{2,\rm neutral}(m)\approx p_0^2(m)$.

\subsection{Non-neutral mutations}
 A non-neutral mutation will necessarily affect the population dynamics; as a result, the fixation probability is no longer strictly independent of $N$.
It is still schematically convenient to formulate the probability of fixation as
\begin{equation}
\label{seq: nonneutral conditional fixation}
    p_{\rm fix}(m | \sigma) \sim  \int d\mathbf{H} p(\mathbf{H} | \sigma) p_{\rm fix}(m | \sigma, \mathbf{H})~,
\end{equation}
where 
\begin{equation}
  p_{\rm fix}(m | \sigma, \mathbf{H}) \sim  1-\exp\left[-r \int_0^{t_{GC}} dt N_\mathbf{H}(t) w (t | \sigma, \mathbf{H})\right]
\end{equation}
Here, we have couched the complexities of the non-neutral case in the fixation probability $w$ of a single mutant arising at frequency $1/N_\textbf{H}$, depending on the evolutionary history $\mathbf{H}$ of the population. Concretely, for a neutral mutation, this is always $w (t | \sigma, \mathbf{H})=1/N_\textbf{H}$, which gives back Eq.~\ref{seq: minimal model}. By contrast, in the non-neutral case, $w$ depends on both the founding clonotype of the lineage $\sigma$ as well as the distribution of genetic backgrounds in the clonotype at the time in which the mutation emerges, which we summarize in a complicated vector of history \textbf{H}. 

Because of this clonotype- and history-dependence, the expected coincident fixation rate of the same mutation at the same site depends on whether the two lineages share the same clonotype.
{\TM Defining
  \begin{equation}
    \mu(m)=\langle p_{\rm fix}(m|\sigma)\rangle_\sigma
  \end{equation}
  as the probability that mutation $m$ fixes in a lineage with a random clonotype $\sigma$, and
  \begin{equation}
    w_2(m)=\langle p_{\rm fix}(m|\sigma)^2\rangle_\sigma
  \end{equation}
  as the probability that $m$ fixes in two independent lineages sharing the same random clonotype $\sigma$, we expect the enrichment statistic
\begin{equation}
    \frac{w_2(m)}{\mu(m)^2}=\frac{\langle p_{\rm fix}(m|\sigma)^2\rangle_\sigma}{\langle p_{\rm fix}(m|\sigma)\rangle_\sigma^2} = 1 + \mathrm{CV}^2_\sigma \big[ p_{\rm fix}(m | \sigma) \big] \geq 1~,
\end{equation}
where we have defined the squared coefficient of variation $\mathrm{CV}_{\sigma}^2 = \mathrm{Var_\sigma}(p_{\rm fix}(m|\sigma))/\langle p_{\rm fix}(m|\sigma)\rangle_\sigma^2$.}
This inequality accounts for the systematic enrichment among nonsynonymous mutations in Fig.~\ref{fig3:landscapes}.

\subsection{On the distinction between fixed and high-frequency mutations}
Most lineages considered in the present study are represented by two sequences. If a mutation is observed in both sequences representing a lineage, the 90\% \DW{CI of the true frequency under a uniform prior is $p\in [0.37, 0.98]$}. Thus, observing the same mutation in $2/2$ sequences does not necessarily imply the mutation is even shared by a majority of cells in the lineage, let alone near fixation. However, we next show that this does not qualitatively limit the applicability of the above analysis in terms of fixation probabilities.

We first note that any lineage represented by multiple sequences is likely of large population size. Concretely, given a total population of memory B cells $N_{\rm mem} \simeq 10^{10}-10^{11}$ in an adult human~\citep{mazzolini2025dynamics}, a lineage represented by $n_{\ell,S} \geq 2$ cells among a sample of size $M_{\rm mem} \ll N_{\rm mem}$ is of typical total size $n_{\ell} \gtrsim N_{\rm mem}  {n_{\ell,S}}/{M_{\rm mem}}$. For the most deeply sequenced individuals considered in this study, about $M_{\rm mem}\sim 10^{6.5}$ memory cells, $n_{\ell} > 10^{3.5}$; in more shallowly sampled individuals, the implied $n_{\ell}$ is still larger. Thus, although the statistical support for the true frequency of the twice-measured mutation is broad, it still implies the mutation has grown to be represented by  thousands of cells.

The key point is then that most of the uncertainty in the fate of a mutation persists while the mutation is at low frequency, i.e. as the mutation began from frequency $1/n_{\ell} \ll 1$. Upon reaching $O(1)$ frequency---whether by drift, selection, or hitchhiking---it has an $O(1)$ probability of then fixing in the lineage, even under fierce clonal competition~\citep{good2012distribution}. \DW{This also justifies our use of the 80\% threshold to define fixation. Our choice allows for high-frequency mutations to augment the coincidence counts, since, from a population genetic perspective, these mutations behave similarly to genuinely fixed mutations.} Misidentification of high frequency mutations as fixed will tend to overestimate the true fixation probability by only $O(1)$ factors, less than the typical order(s) of magnitude distinction between neutral and selected ``fixation" probabilities that we measure.

\section{Mixture model of clonotype-specific selection}
\label{si: mixture model}
This section describes a general mixture model formulation of each mutation's fixation rate in a randomly sampled clonotype, and its relation to the enrichment statistic~(Fig.~\ref{fig3:landscapes}b). We then consider the simplified limit used for inference and cross-validation, as described in the main text~(Fig.~\ref{fig4:mixture model inference}). 

\subsection{Model formulation}

We define $\rho_m(p)=\langle \delta(p-p_{\rm fix}(m|\sigma))\rangle_\sigma$, the distribution of fixation probabilities of mutation $m$ across clonotypes $\sigma$. With this definition, the average fixation probability of $m$ is
\begin{equation}
    \mu(m)=\langle p_{\rm fix}(m|\sigma)\rangle_\sigma = \int dp\, p \rho_m(p),
\end{equation}
and the probability of observing the same fixed mutation $m$ in two independent lineages of clonotypes $\sigma$ and $\sigma'$ is $\mu(m)^2$ if $\sigma\neq \sigma'$, and
\begin{equation}\label{seq: c_sigma,sigma'}
w_2(m)=\langle p_{\rm fix}(m|\sigma)^2\rangle_\sigma=\int dp\, p^2 \rho_m(p)
\end{equation}
if $\sigma'=\sigma$.
\DW{This expression reflects the independent fates of lineages of shared clonotype but arising in different individuals: a mutation's fixation in zero, one, or both lineages depends on two independent Bernoulli trials with the same fixation probability $p(m)$ drawn from $\rho_m(p)$.}

The enrichment plotted in Fig.~\ref{fig3:landscapes} of the main text is then simply an empirical estimate of $w_2(m)/\mu(m)^2$.

Because the enrichment statistic is only a function of the first and second moments of the distribution of fixation probabilities, it is not enough to infer the full distribution. This motivated us to simplify the problem by assuming a bimodal distribution of fixation probabilities:
\begin{equation}\label{seq:delta-fn rho_finite-variance}
  \rho_m(p)=(1-\phi_{\rm b})\delta(p-p_0(m))+\phi_{\rm b} \rho_{m,b}(p),
\end{equation}
where $p_0(m)$ is the neutral fixation probability, $\phi_{\rm b}(m)$ the (small) probability that the mutation is beneficial, and $\rho_{m,b}(p)$ (of support $p>p_0$) the distribution of fixation probabilities {provided the mutation is beneficial (in a fraction $\phi_{\rm b}(m)$ of clonotypes)}. Note that in this Ansatz, we neglect a small class of clonotypes in which the mutation may be strongly deleterious, and thus have negligible chance of fixing, $p_{\rm fix}(m)  \approx 0$. \DW{We later show that such a class must have a self-consistently small effect on our inferences.}

The data is still insufficient to characterize $\rho_{m,b}(p)$. To make progress, we instead make the drastic assumption that it takes a single value $p_{\rm b}(m)$ that is well separated from $p_{0}(m)$ resulting in
\begin{equation}\label{seq:delta-fn mixture}
  \rho_m(p)=(1-\phi)\delta(p-p_0(m))+\phi \delta(p-p_{\rm b}(m))~.
\end{equation}
This simplification is not justified \textit{a priori}, but is necessary if one only has access to shared clonotypes represented by only two independent lineages, as is the case for the majority of our data. Despite the simplifications, one hopes that this simple model still captures some \DW{\textit{characteristic}} value of the fixation probability of beneficial mutations.
\DW{Characteristic may be operationally defined here as the fixation probability that predicts fixation in held-out triplet lineage data (Fig.~\ref{fig4:mixture model inference}d-f). A more statistically motivated interpretation of this characteristic value, in view of the complications ignored by Eq.~\ref{seq:delta-fn mixture}, may be found in \siref{si: model interpretation}.}

\DW{Note that we distinguish here an ``effective" beneficial class size $\phi$ from the true beneficial class size $\phi_{\rm b}$, since in general, the inference procedure described below infers $\phi(m) \leq \phi_{\rm b}(m)$.} 

\subsection{Model inference with lineage pairs of shared clonotype}
For each mutation, Eq.~\ref{seq:delta-fn mixture} describes a three-parameter model, to be inferred from our clonotype-sharing lineage data as follows.
For each clonotype $\sigma$, the mutation $m$ is in the set $\mathcal{A}(\sigma)$ if it is accessible by single point mutations from the annotated germline V gene. For each mutation $m$, we identified the set $\mathcal{G}(m)$ of public clonotypes---represented by $n_\sigma$ expanded lineages in two or more individuals sampled in our data---in which this mutation was accessible, i.e. $\mathcal{G}(m) = \{\sigma | n_\sigma \geq 2, m \in \mathcal{A}(\sigma) \}$. Denote by $k_\sigma(m) \leq  n_\sigma$ the number of lineages among those of shared clonotype $\sigma \in G(m)$ in which the mutation was observed as fixed.
  The log-likelihood of the data given the parameters reads:
\begin{equation} \label{seq: def log-likelihood}    
  \begin{split}
    \mathcal{L} & \equiv \ln P\big(\big\{k_\sigma (m) | \sigma \in \mathcal{G}(m)\big\}~\big|~\phi(m), p_{\rm b}(m), p_{0}(m) \big) \\
    &= \sum_{\sigma} \ln \bigg[  \phi(m) p_{\rm b}(m)^{k_\sigma(m)} (1- p_{\rm b}(m))^{n_\sigma-k_\sigma(m)}  \\
    &+ (1-\phi(m)) p_{0}(m)^{k_\sigma(m)} (1- p_{0}(m))^{n_\sigma-k_\sigma(m)} \bigg] + \textrm{const.},
    \end{split}
\end{equation}
where we have absorbed additive terms that are independent of $(\phi_{\rm b}(m), p_{\rm b}(m), p_{0}(m))$ in the constant. With many clonotypes supported by $n_\sigma \geq 3$ lineages, it is in principle possible to infer the three parameters in $\Theta(m)$ simultaneously. However, the large majority of clonotypes are supported by at most two lineages~(Fig.~\ref{fig1:schematic}c). With only two lineages per clonotypes, the maximum log likelihood is degenerate over $\Theta$.

Rather than rely on small (noisy) numbers of clonotypes with $n_\sigma \geq 3$ to break this degeneracy, we instead used non-productive data to independently fit one of the parameters, $p_{0}(m)$. 
Because a nonproductive sequence is presumed linked to an unknown productive sequence on the other chromosome, the nonproductive lineages experience SHM, drift, and hitchhiking with the productive VDJ sequence, but without direct selection on the mutations on the nonproductive sequence. Thus, the rates of fixation in these non-productive lineages should closely approximate $p_0(m)$. We inferred $p_0(m)$ as the fraction of non-productive lineages, represented by two or more sequences and with the V gene of $m$, where $m$ reached at least 80\% frequency in the lineage.
McCoy~\etal~\citep{mccoy2015quantifying} and Matsen~\etal~\citep{matsen2025sitewise, matsen2026separating} take similar approaches, although here we focus on high-frequency mutations alone.

\subsection{Maximum likelihood estimation is equivalent to method-of-moments for pair data}

Given $p_0$, the unknown parameters $\phi(m)$ and $p_{\rm b}(m)$ may be inferred via the maximum likelihood estimator (MLE):
\begin{align}
\label{seq: argmax log-likelihood}
    (\hat\phi(m),\hat p_{\rm b}(m)) = \underset{\phi(m),p_{\rm b}(m)}{\mathrm{argmax}}~\ln P\big(\big\{k_\sigma (m) | \sigma \in \mathcal{G}(m)\big\}~\big|~\phi(m), p_{\rm b}(m), p_{0}(m) \big).
\end{align}
\DW{For strictly pair data, the MLE has a closed form solution that is equivalent to a method-of-moments approach. Following the notation in the Methods, we write the expected fraction of pairs in which the mutation was observed $k =$ 0, 1, or 2 times as $w_k(\phi(m), p_{\rm b}(m))$, with
\begin{align} 
\label{seq:w_k(phi,p_b)}
	w_k(\phi(m), p_{\rm b}(m)) = \phi(m) \binom{2}{k} p_{\rm b}(m)^{k}(1-p_{\rm b}(m))^{2-k} + (1-\phi(m)) \binom{2}{k} p_0(m)^{k}(1-p_0(m))^{2-k}~,
\end{align}
where the dependence on $p_0(m)$ is kept implicit for clarity. Let $N_k(m)$ equal the number of pairs where the mutation $m$ was observed $k$ times, and $N(m)=N_0(m)+N_1(m)+N_2(m)$. Rewriting the log-likelihood in terms of these parameters, with the constraint $w_0(\phi(m), p_{\rm b}(m)) = 1-w_1(\phi(m), p_{\rm b}(m))-w_2(\phi, p_{\rm b})$, and then maximizing with respect to $w_2$ and $w_1$, one finds
\begin{align}
	\hat{w}_k(m) = \frac{N_k(m)}{N(m)}~,
\label{seq:hat_wi}
\end{align}
for $k=0,1,2$. Define the mean of the mutation's fixation distribution $\mu = \mathbf{E}_{\rho}[p]$, whose unbiased estimator is given by $\hat{\mu} = \hat{w}_2 + \frac{1}{2} \hat{w}_1$. Inverting Eqs.~\ref{seq:w_k(phi,p_b)} gives $\hat{\phi}$ and $\hat{p}_{\rm b}$ as
\begin{subequations}
\label{seq:pb,phi estimators}
\begin{align} 
	& \hat{p}_{\rm b} = \frac{\hat{w}_2 - p_0^2}{\hat{\mu} - p_0} - p_0~, \\
	& \hat{\phi} = \frac{\hat{\mu} -p_0}{\hat{p}_{\rm b}-p_0}~. 
\end{align}
\end{subequations}}
\DW{Equation~\ref{seq:pb,phi estimators} applies under the conditions
\begin{equation}
    p_0 < \hat \mu < 1,\qquad
    \hat{\mu}^2 < \hat w_2 < (1+p_0)\hat \mu-p_0.
\end{equation}
Otherwise, one parameter is constrained to 0 or 1, and the other is fixed via solution of the one-dimensional MLE problem. Defining $z=(1+p_0)\hat\mu-p_0$, the constrained MLE $(\phi^*, p_{\rm b}^*)$ is
\begin{equation}
(\phi^*, p_b^*)=
\begin{cases}
\displaystyle
\left( \hat{\phi}, \hat{p}_{\rm b} \right),
& \hat\mu>p_0~{\rm and}~\hat\mu^2<\hat w_2<z,
\\
(1,\hat\mu),
& \hat\mu\geq p_0~{\rm and}~\hat w_2\leq\hat\mu^2,
\\
\displaystyle
\left(\frac{\hat w_2-p_0^2}{1-p_0^2},1\right),
& \hat w_2\geq p_0^2~{\rm and}~\hat w_2\geq z
\\
(0,\text{unidentified}),
& \hat\mu\leq p_0~{\rm and}~\hat w_2\leq p_0^2.
\end{cases}
\end{equation}
The second and third lines describe the boundary-constrained solutions for some mutations displayed in Fig.~\ref{fig4:mixture model inference}b. The last line describes the case where the statistics are inconsistent with the two-class model where, either because $p_0(m)$ systematically overestimates the true background class for $m$ (those are excluded from Fig.~\ref{fig4:mixture model inference} entirely), or because sampling fluctuations do not support the two-class model (represented as a fraction of the ``neutral" pie chart, the other fraction coming from interior or boundary solutions that failed to reach significance).

It is useful to define the feasible region of $(\phi, p_{\rm b})$ inferences given sampling noise
(in Fig.~\ref{fig4:mixture model inference}b). In particular, we seek a minimum $\tilde{p}_{\rm b}(m; \phi)$ across mutations (Methods) that depends on the dataset size $N_{m}$ and neutral fixation rates $p_0(m)$ across mutations, but integrates out mutations' observed pair statistics (and therefore of $\hat \mu$ and $\hat w_2$) to give a self-consistent lower bound on the feasible region. Its form can be motivated by first noting that, for each mutation, inference of a finite $\hat \phi, \hat p_{\rm b} < 1$ rests on the observation of excess coincidences $N_2$. Then, given $\{N(m), p_0(m)\}$, we require that the observed number of coincidences $N_2$ be inconsistent with the neutral expectation: that is, $N_2$ must be larger than a minimum $n_\alpha$ (for significance level $\alpha$), defined by
\begin{align} 
\label{seq: binom n_alpha}
  n_\alpha(m)=\min\left\{n:
  \Pr\!\left[\operatorname{Bin}\!\left(N(m),p_0(m)^2\right)\ge n\right]\le\alpha
  \right\}.
\end{align}
This is analogous to our bootstrapped likelihood ratio test, but conditioned on $N_2$ alone: instead of dependence on $N_1$, $N_1$ is implicitly fixed by the $\phi$ and $p_b$ consistent with ($N, p_0, N_2$). To wit, Eq~\ref{seq: binom n_alpha} and the definition $w_2 \equiv \phi p_{\rm b}^2 + (1- \phi)p_0^2$ together define  
\begin{align}
\label{seq: tilde pb}
\tilde{p}_{\rm b}(m; \phi, \alpha) = \sqrt{\frac{1}{N(m) \phi} \big(\max\{n_\alpha(m), 2\}  - (1-\phi)p_0^2 N(m)\big)}~,
\end{align}
where we have replaced $n_\alpha(m)$ by $\max\{n_\alpha(m), 2\}$ to account for the fact that Fig.~\ref{fig4:mixture model inference}b requires $N_2 \geq 2$ to be included in the plot. The feasible region curve plotted in Fig.~\ref{fig4:mixture model inference}b is then given by the minimum $\tilde{p}_{\rm b}(m; \phi, \alpha=0.05)$ across all 1259 $m$ considered for each $\phi$. 

\subsection{Induced anticorrelation across mutations}

Figure.~\ref{fig4:mixture model inference}b shows a broad negative linear relation between $\log \hat \phi$ and $\log \hat p_{\rm b}$, of slope $\hat{\beta} {\simeq}-1/2$. This in part reflects ascertainment bias owing to the shape of the feasible region defined in the previous section. In general such a negative correlation can also be consistent with the algebraic relation between $\hat \phi$ and $\hat p_{\rm b}$ in Eqs.~\ref{seq:pb,phi estimators}. In particular, note that in the limit $\hat p_{\rm b} \gg p_0$ for each $m$ (Fig.~\ref{sfig:pben/pneut vs pneut vs mode}a), then $\log \hat \phi \simeq \log (\hat \mu - p_0) - \log \hat p_{\rm b}$. The slope $\beta$ of the linear regression of $y=\log \hat p_{\rm b}$ on $x=\log \hat \phi = \log(\hat \mu - p_0) - \log \hat p_{\rm b} \equiv z-y$ is given as a function of the sampled variances and covariances:
\begin{subequations}
\begin{align} 
  \beta &= \frac{{\rm Cov}(x, y)}{{\rm Var}(x)} = \frac{ {\rm Cov}(z-y, y)}{{\rm Var}(z-y)} \\
  &= \frac{{\rm Cov}(z, y) - {\rm Var}(y)}{{\rm Var}(z) + {\rm Var}(y) - 2{\rm Cov}(z, y)}~,
\end{align}
\end{subequations}
where the second line follows from linearity of covariance. If ${\rm Var}(y) \gtrsim {\rm Cov}(z,y)$, then $\beta$ is dominated by ${\rm Var}(y)$, and merely reflects co-occurrence of $y$ in both the x and y axes. 

To remove such artefactual correlation, it is convenient instead to plot $y$, measuring the direct beneficial fixation rate of mutations,  versus $z$, capturing the effective prevalence ($\sim \phi (p_b-p_0)$) of beneficial fixations of mutations (Fig.~\ref{sfig:pben/pneut vs pneut vs mode}c). Fig.~\ref{sfig:pben/pneut vs pneut vs mode}c evinces a much milder, but still significant, negative correlation,  with non-negligible covariance ${\rm Cov}(z,y) \sim -0.6 {\rm Var}(y)$ (for the 83 mutations with $N_2 \geq 10$), providing some evidence for an empirical anticorrelation beyond the artefact induces by the model.

We next elaborate on the logic of the model cross-validation in the next section, before returning to the algebraic construction Eqs.~\ref{seq:pb,phi estimators} to understand how the parameters assigned by the simple two-component model map to more complicated underlying distributions of fixation probabilities (e.g. Eq.~\ref{seq:delta-fn rho_finite-variance}). Only mutations obeying the interior solution (Eqs.~\ref{seq:pb,phi estimators}) enter in the cross-validation analyses and subsequent analyses.
}

\subsection{Model cross-validation with held-out lineages}

We sought to assess if our inferences meaningfully decomposes the beneficial prevalence $\phi$ from fitness effect $p_{\rm b}$. In particular, Fig.~\ref{fig4:mixture model inference}b indicates \DW{\sout{a broad anticorrelation  between $\phi$ and $p_{\rm b}$}meaningful mutation-to-mutation variation in ($\phi, p_{\rm b}$)}. We wanted to assess whether this reflected real mutation-to-mutation variation in these parameters or instead reflected spurious fitting related to model misspecification or overfitting to noise. To do so, we use shared clonotypes with at least one additional lineage, which were not observed during inference.  Denote $k_\sigma(m)=\{0,1,2\}$ as the number of observations of mutations of amino acid $a$ at site $i$ among two lineages of shared clonotype. The conditional probability of observing the same mutation in a third lineage of the same clonotype, given its observations $k_{\sigma}(m)$ times in the first two lineages, is
\begin{align}
\label{seq:observed in third given pair}
    \mathrm{Prob\big[m~obs.~in~third}| k_\sigma(m) \big] &= \frac{ \langle p^{1+k}(1-p)^{2-k} \rangle}{ \langle p^{k}(1-p)^{2-k} \rangle } \\
\label{seq:observed in third given pair two-mixture}
     &= \frac{ \phi p_{\rm b}^{1+k_\sigma(m)}(1-p_{\rm b})^{2- k_\sigma(m)} + (1-\phi)p_0^{1+ k_\sigma(m)}(1-p_0)^{2- k_\sigma(m)}}{\phi p_{\rm b}^{ k_\sigma(m)}(1-p_{\rm b})^{2- k_\sigma(m)} +(1-\phi)p_0^{ k_\sigma(m)}(1-p_0)^{2- k_\sigma(m)} }~,
\end{align}
where the second equality follows under the assumption of the two-point mixture model (Eq.~\ref{seq:delta-fn mixture}).
The observed rates of fixation in the third lineage for each $m$, conditioned on the observations $k_\sigma(m)$, are shown in~Fig.~\ref{fig4:mixture model inference}d-f. The contribution of beneficial mutations plotted in the inset is given by:
\begin{equation}
g_{\rm ben}(m|k_\sigma(m))=\frac{ \phi p_{\rm b}^{1+k_\sigma(m)}(1-p_{\rm b})^{2- k_\sigma(m)}}{ \phi p_{\rm b}^{1+k_\sigma(m)}(1-p_{\rm b})^{2- k_\sigma(m)} + (1-\phi)p_0^{1+ k_\sigma(m)}(1-p_0)^{2- k_\sigma(m)}}\,.
\end{equation}
    In the limit $p_0(m)<p_{\rm b}(m)\ll 1$, Eq.~\ref{seq:observed in third given pair two-mixture} becomes
\begin{align}
   \mathrm{Prob}\big[m\textrm{ observed in third}| k_\sigma(m) \big]  \simeq p_{\rm b}\frac{1+\frac{1-\phi}{\phi}\frac{p_0^{1+ k_\sigma(m)}}{p_{\rm b}^{1+ k_\sigma(m)}}}{1+\frac{1-\phi}{\phi}\frac{p_0^{ k_\sigma(m)}}{p_{\rm b}^{ k_\sigma(m)}}}\,. 
\end{align}
In particular, provided that $(p_{\rm b}/p_0)^{1+k_\sigma(m)} \gg (1-\phi)/\phi$, we have $g_{\rm ben}\approx 1$ and
\begin{align}
\mathrm{Prob}\big[m\textrm{ observed in third}| k_\sigma(m) \big] \simeq p_{\rm b}(m)~.
\end{align}
In this limit, this statistic does not depend on $\phi(m)$, and thus directly assesses whether the model inference meaningfully decouples beneficial prevalence from beneficial fitness effect. The agreement between prediction and observation in this limit is evident when coarse-graining across mutations of similar nominal $p_{\rm b}$. By contrast, predictions made using an averaged point estimate of $\phi(m) = 0.02$ and $p_{\rm b}(m) = 30\cdot p_{0}(m)$ for all $m$ provide a poor fit to data (Fig.~\ref{sfig:pben/pneut vs pneut vs mode}). Thus, despite its simplicity, \DW{the model quantifies meaningful variation across mutations in their characteristic strength and prevalence, where characteristic is a notion that we make precise in the next section}.

\DW{
\section{Consequences of and bounds on model misspecification \label{si: model interpretation}}

We would like to better interpret the inferred fits, in view of more complicated structure in $\rho_{m}(p)$ than that which the minimal two-point mixture of delta functions admits. For brevity in what follows, we drop the mutation index $m$. 

\subsection{A broad distribution of beneficial fixation rates $\rho_{b}$}

We first consider the effects of a beneficial distribution $\rho_{\rm b}$ of mean $\mu_{\rm b}$ and finite variance $\mathrm{Var}_{\sigma}(p|{\rm b})\equiv \mathrm{Var}(\rho_{m,\rm b})$~(Eq.~\ref{seq:delta-fn rho_finite-variance}), while continuing to assume that the background distribution is a delta function centered on $p_0$. In this case, the first and second moments are
\begin{subequations}
\begin{align}
	& \mu = (1-\phi_{\rm b}) p_0 + \phi_{\rm b} \mu_{\rm b} \\
	& w_2 = \langle p^2 \rangle_\rho = (1-\phi_{\rm b}) p_0^2 + \phi_{\rm b} \mu_{\rm b}^2(1 + \mathrm{CV}_{\rm b}^2)~,
\end{align} 
\end{subequations}
where $CV_{\rm b}^2 =\mathrm{Var}_\sigma(p|{\rm b}) / \mu_{\rm b}^2$. Note that these are the true population first and second moments. The MLE implicitly calculates the unbiased estimates of the moments, $\mathbf{E}[\hat{\mu}] = \mu, \mathbf{E}[\hat{w_2}] = w_2$. Plugging into Eq.~\ref{seq:pb,phi estimators}, one obtains
\begin{subequations} \label{seq:pb,phi with variance}
\begin{align} 
	& \mathbf{E}[\hat{p}_{\rm b}-p_0] =  (\mu_{\rm b}-p_0)(1 + \frac{\mathrm{Var}_\sigma(p|{\rm b})}{(\mu_{\rm b}-p_0)^2}) + O(1/N)~ \\
	& \mathbf{E}[\hat{\phi}] = \phi_{\rm b} \cdot (1 + \frac{\mathrm{Var}_\sigma(p|{\rm b})}{(\mu_{\rm b}-p_0)^2})^{-1} + O(1/N)~.
\end{align}
\end{subequations}
The inferred $\hat{p}_{\rm b}$ is upward biased, relative to its conditional mean $\mu_{\rm b}$, while $\hat{\phi}$ is downward biased to compensate. In particular, the expectation Eq.~\ref{seq:pb,phi with variance}a of the inference is related to the ratio of the $p_0$-centered first and second moments of $\rho_{\rm b}$:
\begin{align}
	\mathbf{E}[\hat{p}_{\rm b}-p_0] \overset{N\rightarrow\infty}{=} \frac{\langle (p-p_0)^2 \rangle_{\rho_{\rm b}}}{\mu_{\rm b}-p_0}
\end{align}
Thus the characteristic effect size in \siref{si: mixture model} is the variance-inflated mean of $\rho_{\rm b}$. Strong variation in $\rho_{\rm b}$ can drive the inference above its typical value defined by the mean sense.

However, $\mathbf{E}[\hat{p}_{\rm b}-p_0]$ is the typical value---in the mean sense---of the distribution
\begin{align}
	q_{b}(p) \equiv \frac{(p-p_0) \rho_{b}(p)}{(\mu_{\rm b}-p_0)},
\end{align}
which is the excess fixation-weighted distribution among beneficial clonotypes. With this definition, we have $\langle p-p_0\rangle_q \simeq \mathbf{E}[\hat{p}_{\rm b}-p_0]$ (in the limit $N\rightarrow \infty$). Thus, the inferred $\hat{p}_{\rm b}$ corresponds to the mean of the excess fixation-weighted mean distribution across beneficial clonotypes. Intuitively, unlike $\rho_{\rm b}$, this distribution fairly represents the clonotypes in which the mutation is actually found fixed post affinity maturation, or as seen by an experimentalist sampling clonotypes with the mutation. Conversely, this estimator tends to downweight any nearly neutral mutations sitting just above $p_0$. Thus, in the absence of a deleterious class, the estimator calculates the mean of another natural, if less immediately obvious, distribution of fixation rates.

\subsection{Perturbations from the neutral background distribution $p_0$}
We next consider the effects of distortions in the neutral background distribution, centered on $p_0$. We may generically include this distortion as 
\begin{align}
	\rho_m(p) = (1-\phi - \Delta) \delta(p-p_0) + \phi \delta(p-p_{\rm b}) + \Delta \rho_\Delta(p)~,
\end{align}
where for clarity we have re-summarized the positive background as a delta-function (with appropriately biased $\phi$ and $p_{\rm b}$). As an example one may take $\rho_\Delta(p) = \delta(p)$ to represent clonotypes for which the mutation is sufficiently deleterious to never fix, given the dataset size (e.g. $p \lesssim 1/N$). Alternatively, $\rho_\Delta(p) = \delta(p-\alpha p_0)$ and $\Delta =1-\phi$ if $p_0$ was misestimated from the non-productive lineages. In general, as above, it is useful to summarize $\rho_\Delta(p)$ in terms of its mean $\mu_\Delta$ and variance $\sigma_\Delta^2$. With some algebra, one finds that the MLE infers
\begin{subequations} \label{seq: deleterious class distortion}
\begin{align}
	& \hat{p}_{\rm b} = \frac{p_{\rm b}}{1+\epsilon} \left[ 1 + \epsilon \cdot \frac{1}{p_{\rm b}} \left(\mu_\Delta + \frac{\sigma_\Delta^2}{\mu_\Delta-p_0}\right) \right] \\
	& \hat{\phi} = \phi \frac{(1+\epsilon)^2}{1+\epsilon \frac{\mu_\Delta-p_0}{p_{\rm b} - p_0}\left(1+ \frac{\sigma_\Delta^2}{(\mu_\Delta-p_0)^2}\right)}
\end{align}
\end{subequations}
where
\begin{align}
\label{seq:epsilon def}
	\epsilon \equiv \frac{\Delta(\mu_\Delta - p_0)}{\phi(p_{\rm b} - p_0)}.
\end{align}

While the above applies equally to the context of a globally underestimated background class (e.g. $\mu_\Delta>0$), we now consider the presence of a predominantly deleterious class ($\mu_\Delta < p_0$). 
For such a class, $\epsilon < 0$, and the inference evidently inflates $\hat{p}_{\rm b}$ and deflates $\hat{\phi}$ even more so, such that $\hat{p}_{\rm b} \hat{\phi} \simeq p_{\rm b} \phi (1+\epsilon) < p_{\rm b} \phi$ (for $p_{\rm b} \gg p_0$). Note that for a sufficiently large and deleterious class, $\epsilon$ grows below $-1$: this corresponds to the case where the observed average fixation rate $\mu$ is less than $p_0$. However, such cases are not distorted but discarded from our fit. In the next subsection, we show how lesser distortions Eq.~\ref{seq: deleterious class distortion} combine with variation in the beneficial distribution $\rho_{\rm b}$ to set the scale of the third lineage prediction error in Fig.~\ref{fig4:mixture model inference}d-f.

\subsection{Third lineage predictions bound the scale of model misspecification}

In the absence of a deleterious class, the third lineage prediction error (Fig.~\ref{fig4:mixture model inference}d-f) directly probes the variation in the beneficial distribution $q_{\rm b}$. 
Note that Eq.~\ref{seq:observed in third given pair} involves the third moment of the underlying fixation rate distribution, whereas the model only learns (fits) the first two moments. Concretely, the scale of disagreement between prediction and observation in Fig.~\ref{fig4:mixture model inference}d-f can be turned into a bound on the typical per-mutation size of $\mathrm{CV}^2_q$. For example, when $k=2$ Eq.~\ref{seq:observed in third given pair} gives
\begin{align}
\label{seq:t2 true}
	 \mathrm{Prob}[\mathrm{obs.~in~3rd} | k=2]  = T_{2} = p_0 + \frac{\phi_{\rm b}(\mu_{\rm b}-p_0)}{w_2} \langle p^2 \rangle_q
\end{align}
Meanwhile, the prediction $\hat{T}_2$ of the two point model is given by
\begin{align}
\label{seq:t2 predicted}
	 \hat{T}_{2} = p_0 + \frac{\hat{\phi}(\hat{p}_{\rm b} - p_0)}{w_2} \hat{p}_{\rm b}^2~.
\end{align}
Similar relations may be derived for $k=0$ and 1, and in turn these relations can be related to the coefficient of variation of $q$
\begin{align}
\label{seq:CV^2_q from third lineage prediction error}
	\mathrm{CV}^2_q = \frac{T_k - \hat{T}_k}{\langle p \rangle_q^2(\mu - p_0)} \cdot 
    \begin{cases}
	(1-2\mu + w_2), & k=0 \\
	(w_2-\mu), & k=1  \\
    w_2, &k=2
	\end{cases}~,
\end{align}
where 
\begin{align}
	\mathrm{CV}^2_q \equiv \frac{Var_q(p)}{\langle p \rangle_q^2}~.
\end{align}
Hence the scale of prediction error is directly related to the variation in the fixation-weighted distribution $q_{\rm b}$. The good agreement between prediction and observation among coarse-grained mutation bins in Fig.~\ref{fig4:mixture model inference}d-f would imply, in the typical absence of a substantial deleterious class, a systematically modest $\mathrm{CV}_q^2$. 

However, by inflating the inferred $\hat{p}_{\rm b}$ (Eq.~\ref{seq: deleterious class distortion}a), the presence of a deleterious class could, in principle, counteract the predicted underestimation of 
the third lineage test under substantial variation $\mathrm{\mathrm{CV}}^2_q$ (Eq.~\ref{seq:CV^2_q from third lineage prediction error}). In particular, this mechanism could explain the good coarse-grained agreement between prediction and observation in Fig.~\ref{fig4:mixture model inference}f: mutations whose errors are dominated by unmodeled beneficial variation in $q_{\rm b}$, driving underprediction, may be canceled out by mutations whose prediction errors are dominated by unmodeled deleterious variation, driving overprediction. 

On the other hand, we still observe good quantitative agreement of prediction and observation of the third moment between the best measured \emph{individual} mutations, namely those with the most $k=2$-conditioned triplets to observe (Fig.~\ref{sfig:concealment of beneficial}a). The uniformly consistent agreement of observation with expectation for individual mutations is hard to explain by consistent cancellation of deleterious distortions with beneficial variation, since one generically expects that mutations vary idiosyncratically in their deleterious and beneficial fixation distributions. This provides some indirect evidence that our inferred characteristic $\hat{p}_{\rm b} = \langle p \rangle_q$ is a reasonable (order of magnitude) approximation of the typical $p$ in the excess fixation-weighted distribution $q$. 

Supposing, nonetheless, that there is such a conspiracy of per-mutation cancellations, we may determine how large the deleterious distortion in each mutation would have to be to mask substantially large variation in beneficial effects. Taking $T_2 = \hat{T}_2$ (up to measurement noise), then combining the above results (Eqs.~\ref{seq: deleterious class distortion}-\ref{seq:t2 predicted}), one finds that the deleterious contribution to a mutation's fixation distribution must satisfy $\Delta \geq \Delta_{\rm min}$, where 
\begin{align}
	\Delta_{\rm min} = \mathrm{CV}^2_{q,{\rm min}} \cdot \frac{\mu - p_{0}}{p_{0} -  \mu_\Delta}~,
\end{align}
for a given minimum $\mathrm{CV}^2_{q,{\rm min}}$. Note that this applies for an \emph{arbitrary} distribution of deleterious fixation rates $\rho_\Delta(p)$ with mean $\mu_\Delta$. Fig.~\ref{sfig:concealment of beneficial}b plots, for each well-measured mutation, the requisite $\Delta_{\rm min}$ vs $\mu_\Delta$, for a $\mathrm{CV}^2_{q,{\rm min}} = 0.25$, corresponding to moderate beneficial fixation dispersion. For instance, we find that nearly every mutation would have to be strictly lethal ($p = 0$) in $\gtrsim 10\%$ of clonotypes to mask a moderate $\mathrm{CV}^2_q \sim 0.25$. Such a population fraction is substantial, although not implausible. Conversely, these results show that our inferences and predictions are not unreasonably sensitive to a vanishing fraction of deleterious or lethal backgrounds.

We briefly consider another possible misspecification of the background class: when the true (nearly) neutral class is centered at some $p_0' > p_0$. This could plausibly arise due to sampling noise in the non-productive lineages used to estimate $p_0$. Concretely, consider $p_0 < \mu_\Delta \lesssim 2 p_0$. 
An increasingly large nearly-neutral class $\Delta$ will be absorbed by the nominally adaptive class ($\phi, p_{\rm b}$), increasing $\phi$ and decreasing $p_{\rm b}$. Thus, this distortion is conservative, whereas frequently large $p_{\rm b}$ fits across mutations accessible by our data suggest this source of misparameterization does not dominate the model. More importantly, such a distortion--combined with beneficial class variation-- should tend to systematically underestimate fixation in a third lineage, conditioned on coincident observation in the first two lineages ($k=2$). This is not observed in the data (Fig.~\ref{fig4:mixture model inference}). Thus, we conclude that this source of misparameterization is not systematically dominant.
}

\clearpage
\begin{figure}
    \centering
    \includegraphics[width=1.0\linewidth]{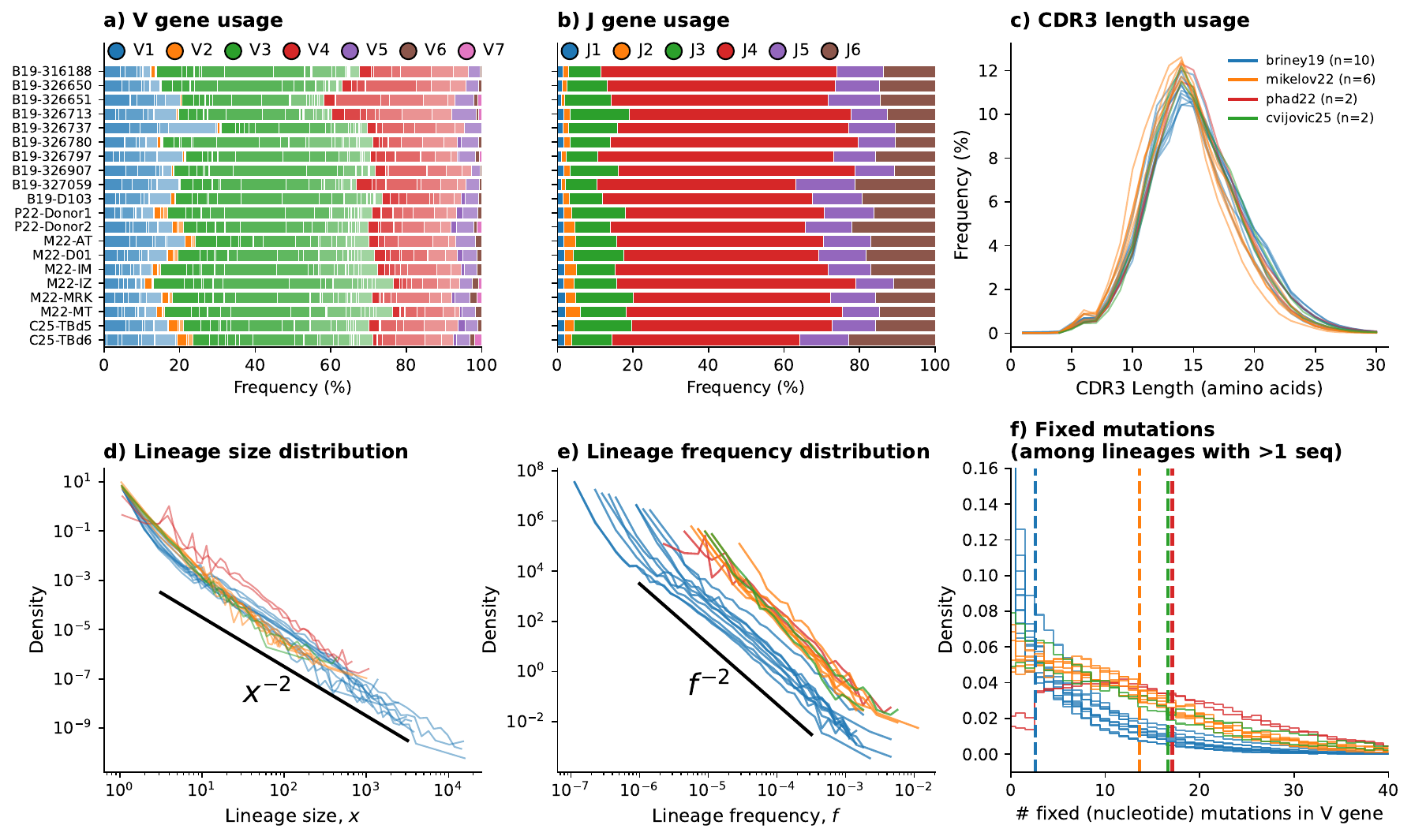}
    \caption{\textbf{Consistent repertoire statistics across subjects and datasets.} Panels (\textbf{a},\textbf{b}) show V and J gene usage as stacked bars for each subject, among lineages constructed from filtered sequences. In (\textbf{a}), hue indicates V gene family and shade indicates the V gene within that family. In the distribution panels, curves are shaded by dataset (blue for B19, orange for M22, red for P22, and green for C25). The panels show (\textbf{c}) CDR3 length, lineage size by (\textbf{d}) raw count and (\textbf{e}) count normalized by sampling depth, and (\textbf{f}) the number of fixed nucleotide mutations ($\geq80\%$ frequency) in the V gene among lineages with at least two sequences.}
    \label{sfig:subject repertoire statistics}
\end{figure}

\clearpage
\begin{figure}
    \centering
    \includegraphics[width=0.9\linewidth]{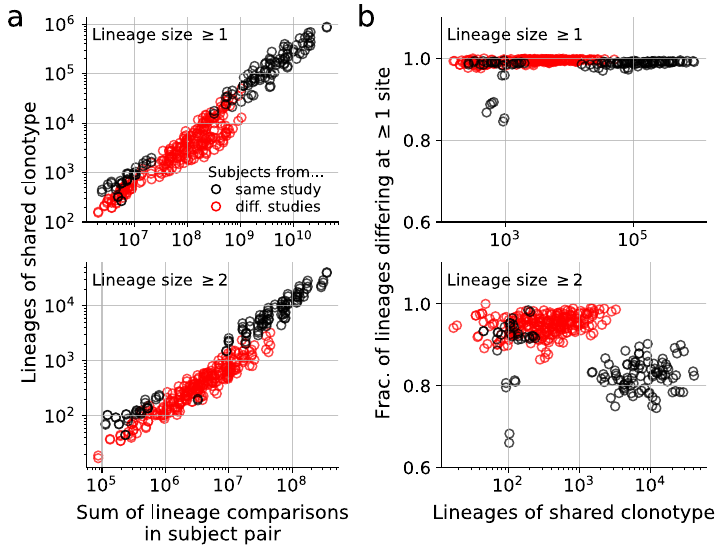}
    \caption{ \textbf{Comparable clonotype sharing rates within and between studies.} (\textbf{a}) Each circle represents a pair of subjects ($s_1$, $s_2$). Plotted on the y-axis is the number of lineages in $s_1$ sharing a clonotype with at least one lineage in $s_2$. The x-axis equals the product of the numbers of $s_1$ and $s_2$ lineages sharing the same VJ$\ell$ (V gene, J gene, and CDR3 length), summed over all VJ$\ell$ groups. (\textbf{b}) For each subject pair, the fraction of lineages of shared clonotype with at least one ``fixed" difference (differing by at least one nucleotide among sites with consensus nucleotides in both lineages).}
    \label{sfig:sharing vs pairwise comparisons}
\end{figure}

\clearpage
\begin{figure}
    \centering
    \includegraphics[width=0.9\linewidth]{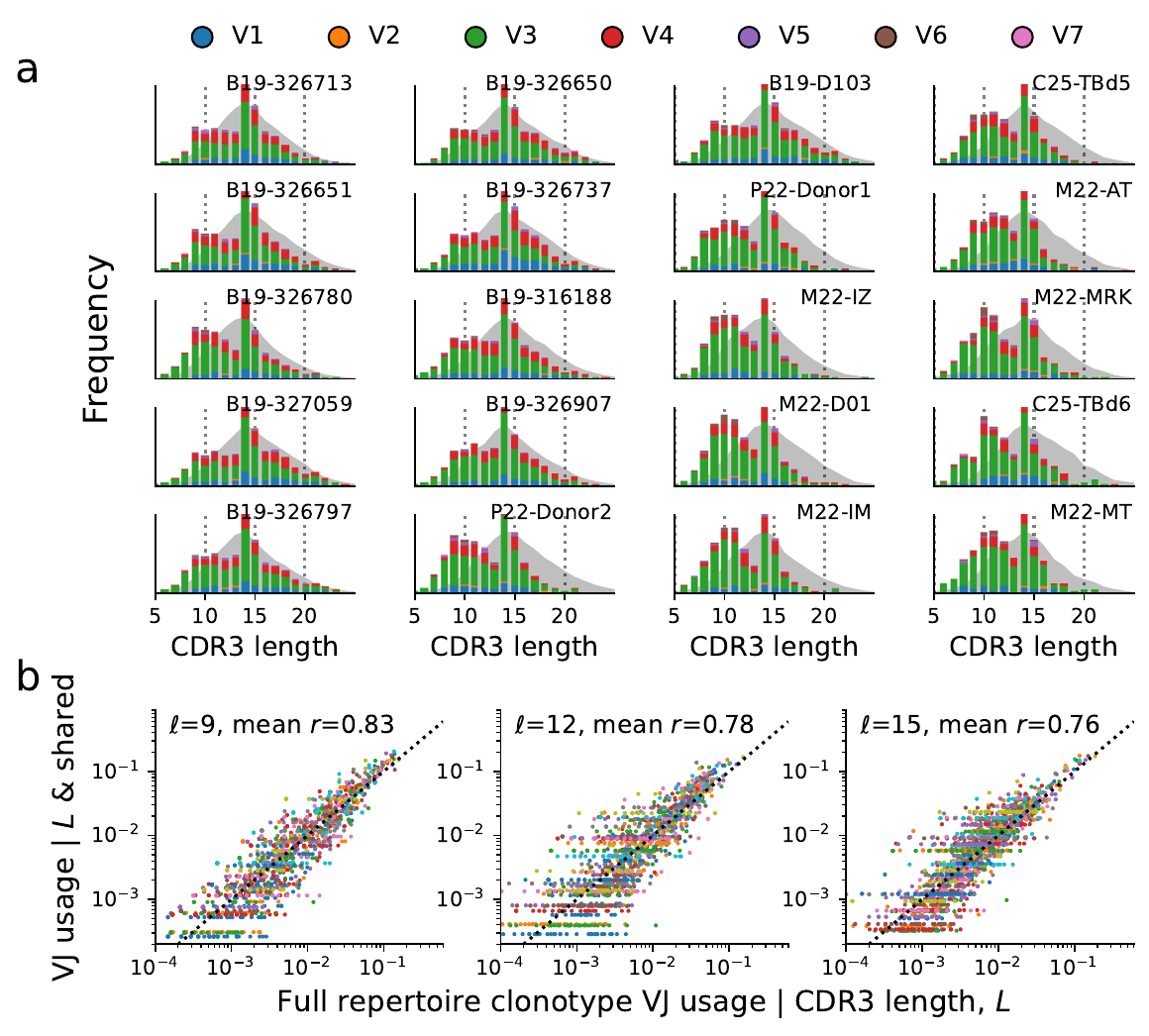}
    \caption{\textbf{VJ$\ell$ usage of shared clonotypes.} (\textbf{a}) For every subject, the distribution of CDR3 lengths of lineages sharing a clonotype with at least one other subject (barplots) against that subject's full distribution of CDR3 lengths (grey). The color of each bar denotes the V gene family. (\textbf{b}) For three characteristic CDR3 lengths, the frequency of a V-J pair among all lineages of shared clonotype, conditioned on the CDR3 length, versus that frequency in the full repertoire, also conditioned on CDR3 length. Colors corresponds to different subjects, and the mean $r$ reports the unweighted average across subjects of the Pearson correlation between the logarithms of the frequencies.}
    \label{sfig:vj3 shared usage}
\end{figure}

\clearpage
\begin{figure}
    \centering
    \includegraphics[width=0.85\linewidth]{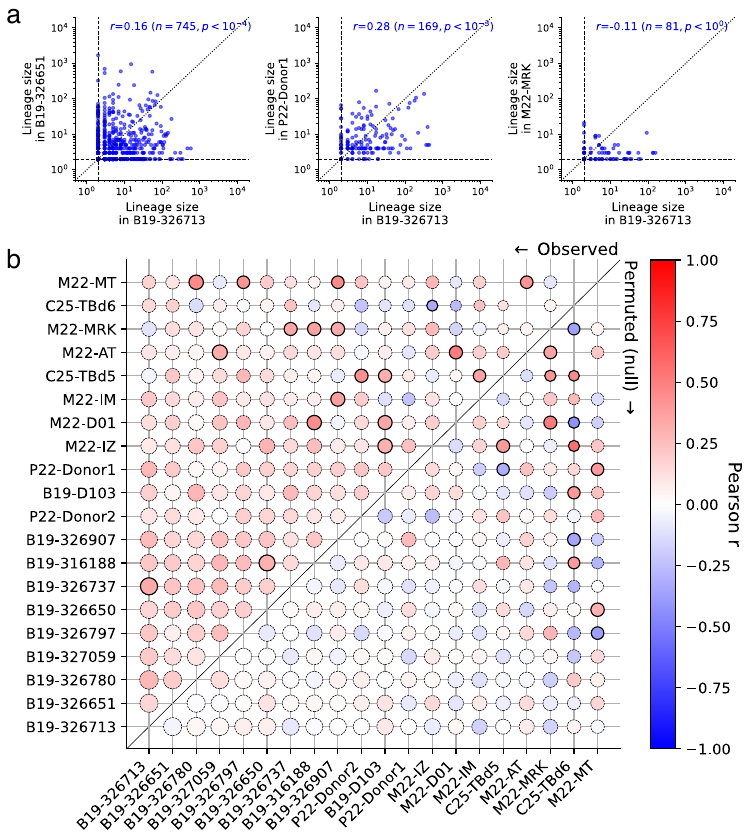}
    \caption{\textbf{Weak but systematic correlations in lineage abundance among inter-subject shared clonotypes.} (\textbf{a}) Example correlation plots of lineage sizes (number of deduplicated sequences/cells) among pairs of lineages of shared clonotype, for three pairs of subjects. Only clonotypes with lineages represented by $\geq 2$ (deduplicated) sequences/cells in both subjects are shown. 
    (\textbf{b}) Pearson correlations of log(lineage abundance) among shared clonotypes between every pair of subjects. Subject pairs with fewer than 10 such clonotypes are excluded. The size of each circle is scaled by the (log) number of clonotypes included in the pair~(10-1087). Bolded circles indicate Pearson correlations $r{>}0.30$. The observed correlations (upper left triangle) are systematically positively correlated, relative to the null expectation (bottom right triangle)~(Methods). Fourteen inter-dataset pairs have $r>0.30$, compared to $5.3{\pm}4.3$ (mean $\pm$ 2 standard deviations), estimated from 100 null permutations.
    }
    \label{sfig:abundance correlations}
\end{figure}

\clearpage
\begin{figure}
    \centering
    \includegraphics[width=1.0\linewidth]{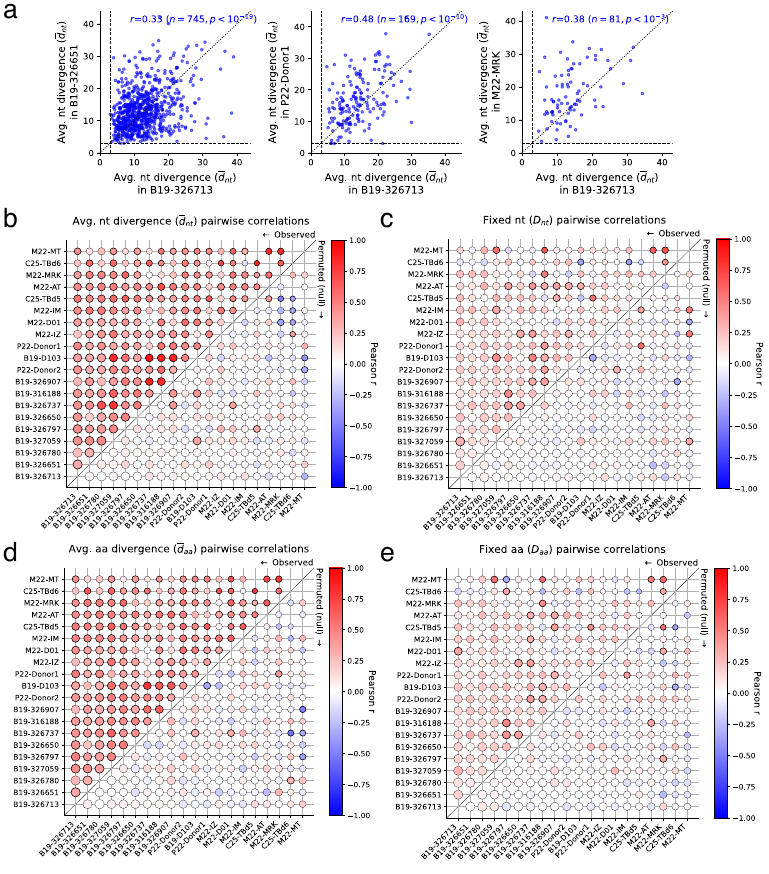}
    \caption{\textbf{Moderate systematic correlations in rates of molecular evolution across subjects.} (\textbf{a}) 
    Example correlation plots of average nucleotide divergence (number of mutations) from germline, in the non-CDR3 V and J regions, among pairs of lineages of shared clonotype. Only clonotypes with lineages represented by ${\geq}2$ sequences or cells in both subjects are shown.
    (\textbf{b-e}) Analogue of Fig.~\ref{sfig:abundance correlations}b, for four other lineage measures: (\textbf{b}) average number of nucleotide mutations $\bar{d}_{\rm nt}$ in the lineage and    (\textbf{c}) number of fixed nucleotide mutations $D_{\rm nt}$; (\textbf{d})average number of amino acid mutations $\bar{d}_{\rm aa}$ in the lineage (\textbf{e}) and number of fixed amino acid mutations $D_{\rm aa}$ (Methods).
    The observed correlations (upper left triangle) are systematically positively correlated, relative to the null expectation (bottom right triangle; see Methods). Substantial positive correlations ($r{>}0.30$ subject pairs in bold) are consistently enriched in the observed data versus the permuted null.
    }
    \label{sfig:macroevo correlations}
\end{figure}

\clearpage
\begin{figure}
    \centering
    \includegraphics[width=1.0\linewidth]{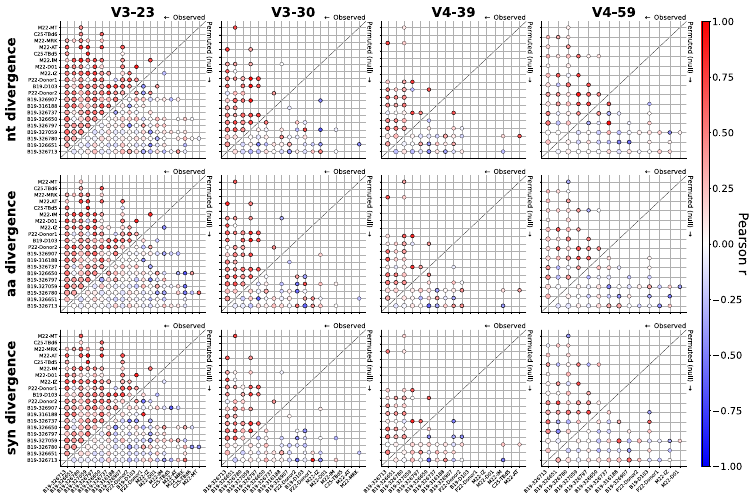}
    \caption{ \textbf{Analogue of Fig.~\ref{sfig:macroevo correlations}, conditioned on clonotypes sharing the same V gene.} Two common genes per family in the V3 and V4 families~(Fig.~\ref{sfig:subject repertoire statistics}) are shown. Consistent correlations within groups of shared clonotypes with the same V gene rule out that the repertoire-wide correlations~(Fig.~\ref{sfig:macroevo correlations}) are driven by variation in mutability under SHM for different V genes.
    }
    \label{sfig:gene specific macroevo correlations}
\end{figure}

\clearpage
\begin{figure}
    \centering
    \includegraphics[width=1.0\linewidth]{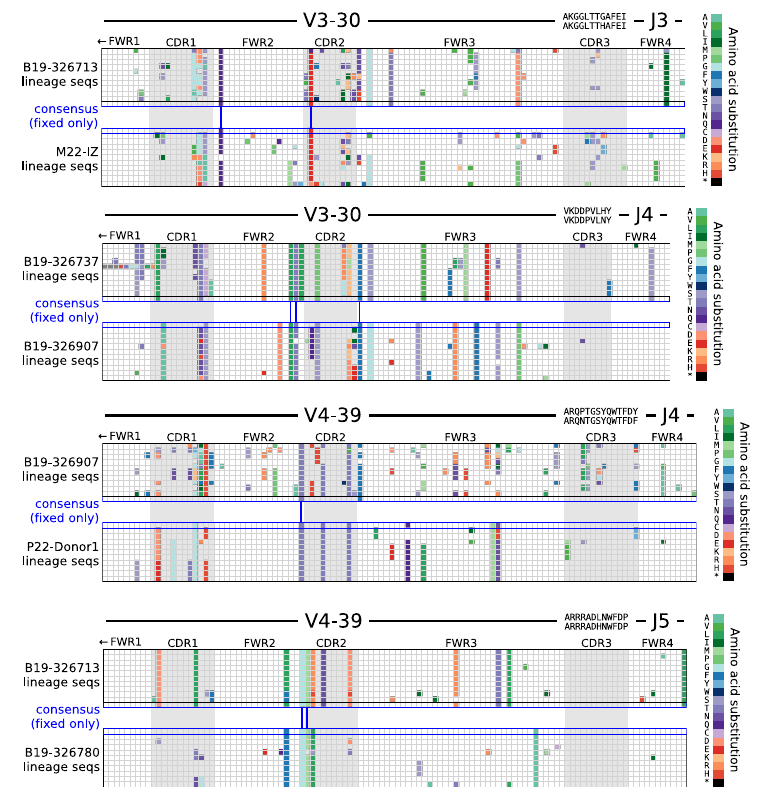}
    \caption{ \textbf{Analogue of Fig.~\ref{fig2:correlations}a.} Four additional example amino acid alignments of lineage pairs of shared clonotype. Examples were randomly chosen among V3-30 and V4-39 lineage pairs with at least ten sequences per lineage and at least one coincident fixation.}
    \label{sfig:example amino acid alignments}
\end{figure}

\clearpage
\begin{figure}
    \centering
    \includegraphics[width=1.0\linewidth]{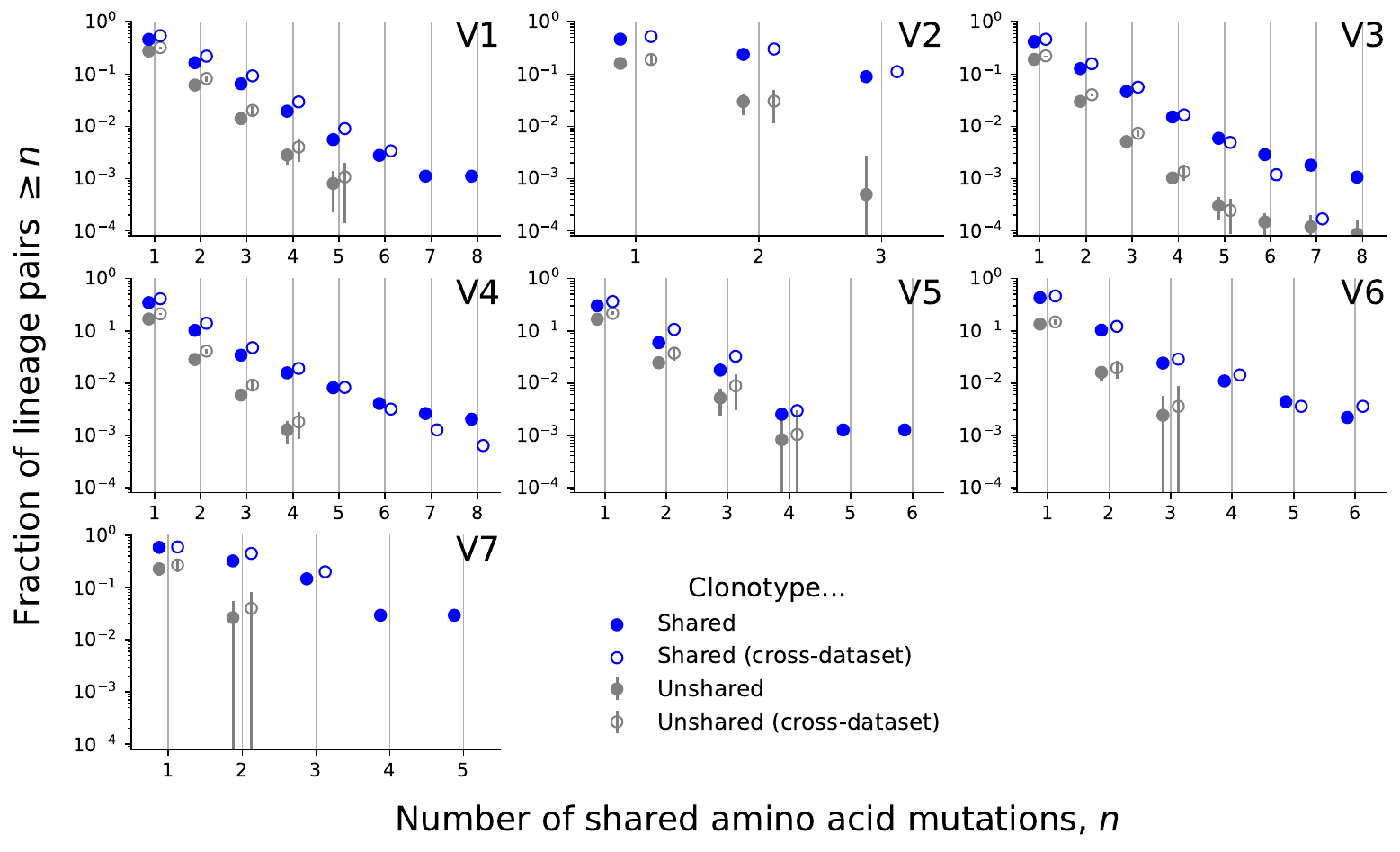}
    \caption{ \textbf{Analogue of Fig.~\ref{fig2:correlations}c, splitting clonotypes by V family.} Blue points represent lineage pairs of shared clonotype, and grey points represent lineage pairs of unshared clonotype. Complementary cumulative functions were also calculated using only lineage pairs deriving from subjects of different datasets, which eliminates any role for contamination. Because most intra-dataset pairs derive from B19, which are less mutated on average~(Fig.~\ref{sfig:subject repertoire statistics}f), the overall rate of coincident fixations is \textit{higher} among lineages of inter-dataset pair. Crucially, however, the fold-enrichment of coincident fixation in pairs of shared clonotype, relative to those of unshared clonotype, is quantitatively consistent whether subjects derive from the same or different datasets.}
    \label{sfig:vfam coincidence survival functions}
\end{figure}

\clearpage
\begin{figure}
    \centering
    \includegraphics[width=0.8\linewidth]{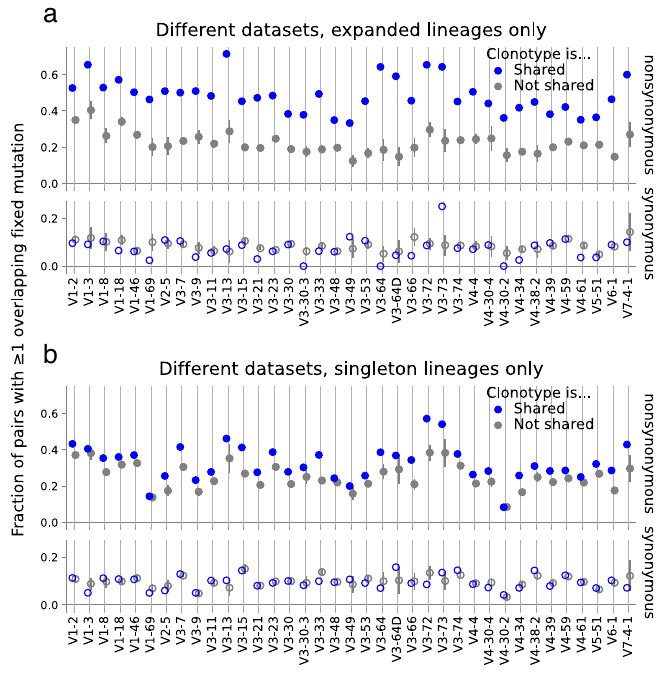}
    \caption{ \textbf{Analogue of Fig.~\ref{fig2:correlations}d, considering only lineage pairs from different datasets.} (\textbf{a}) Coincidence enrichments conditioned on clonotype-matched lineages (represented by two or more sequences) deriving from subjects from different datasets. As in Fig.~\ref{sfig:vfam coincidence survival functions}, the scale of the enrichment is quantitatively similar to the unconditioned enrichment (Fig.~\ref{fig2:correlations}d), implying minimal impact from intra-dataset contamination. (\textbf{b}) Sampling only clonotype-matched pairs where each lineage is represented by a singleton sequence significantly diminishes the signal-to-noise null ratio.
    }
    \label{sfig:vgene panel different datasets}
\end{figure}

\clearpage
\begin{figure}
    \centering
    \includegraphics[width=0.95\linewidth]{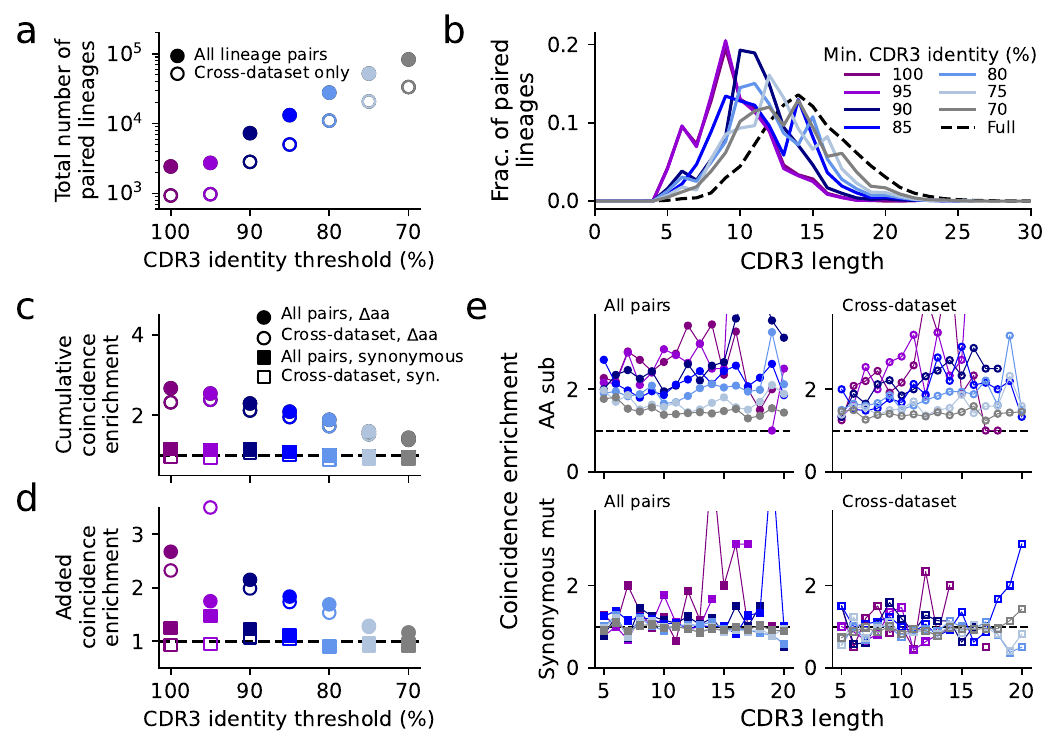}
    \caption{\DW{\textbf{Analysis of clonotype sharing across minimum CDR3 identity thresholds.} (\textbf{a}) Number of lineage pairs, aggregated across all subjects, of shared clonotype at varying CDR3 percent identity thresholds. Each lineage is represented by exactly one pair. By construction, the set of shared-clonotype lineages at CDR3 identity threshold $d_1$ is a subset of the shared-clonotype lineages at any smaller CDR3 identity threshold $d_2<d_1$. Pairs with lineages deriving from different datasets (open circles) represent ${\sim}1/3$ of all lineage pairs (filled circles). (\textbf{b}) At reduced CDR3 identity thresholds, the distribution of CDR3 lengths among pairs shifts right, approaching the distribution among all lineages in a typical subject (here, B19-326713). (\textbf{c}) Related to Fig.~\ref{fig2:correlations}c, the enrichment in the rate of coincident fixed mutation (i.e. sharing at least one fixed mutation) in shared-clonotype pairs versus unrelated-clonotype pairs, as a function of CDR3 identity threshold. Circles represent amino-acid mutations, while squares represent synonymous nucleotide mutations, which show no enrichment at any threshold.
      (\textbf{d}) The fixed amino acid mutation coincidence enrichment among the additional lineage pairs at $d_i$ not represented by $d_{i-1}$ decays to zero at $d_i \sim 70\%$, indicating decorrelated adaptive landscapes at small CDR3 identities and consequently reduced cumulative enrichment in \textbf{c}. (\textbf{e}) Fixed coincidence enrichment as a function of CDR3 length, for both amino acid substitutions (top row) and synonymous mutations (bottom).}
    }
    \label{sfig:cdr3 distance threshold analysis}
\end{figure}

\clearpage
\begin{figure}
    \centering
    \includegraphics[width=1.0\linewidth]{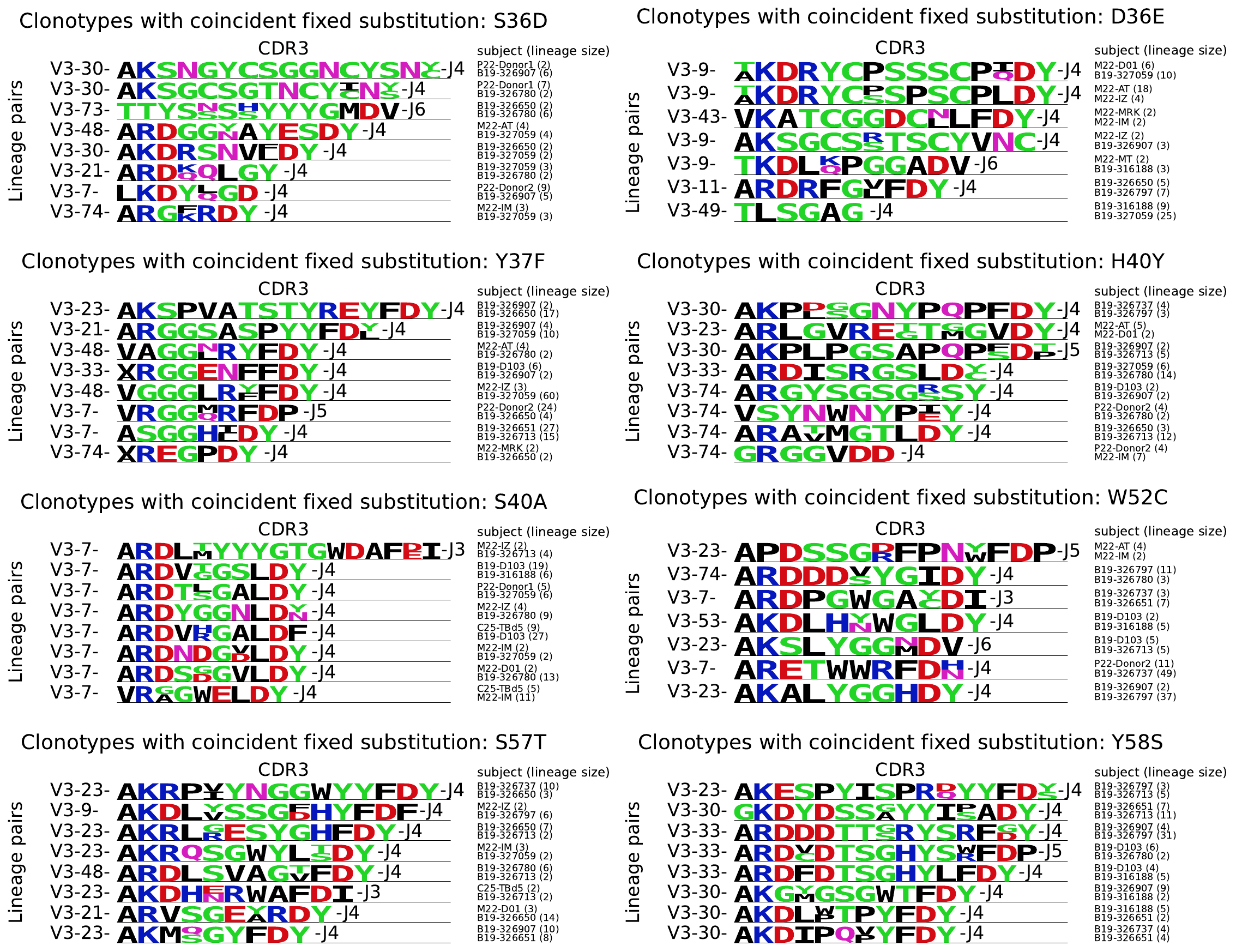}
    \caption{ \textbf{Analogue of Fig.~\ref{fig3:landscapes}c.} Additional example substitutions, and random clonotypes in which fixations of these mutations were observed in both lineages. Across examples, the coincidence enrichment over the null ranges from $X\simeq2.5-11$, implying that a fraction $X/(X+1) > 70\% $ of coincidences were driven by convergent selection.}
    \label{sfig:more logo plots}
\end{figure}

\clearpage
\begin{figure}
    \centering
    \includegraphics[width=1.0\linewidth]{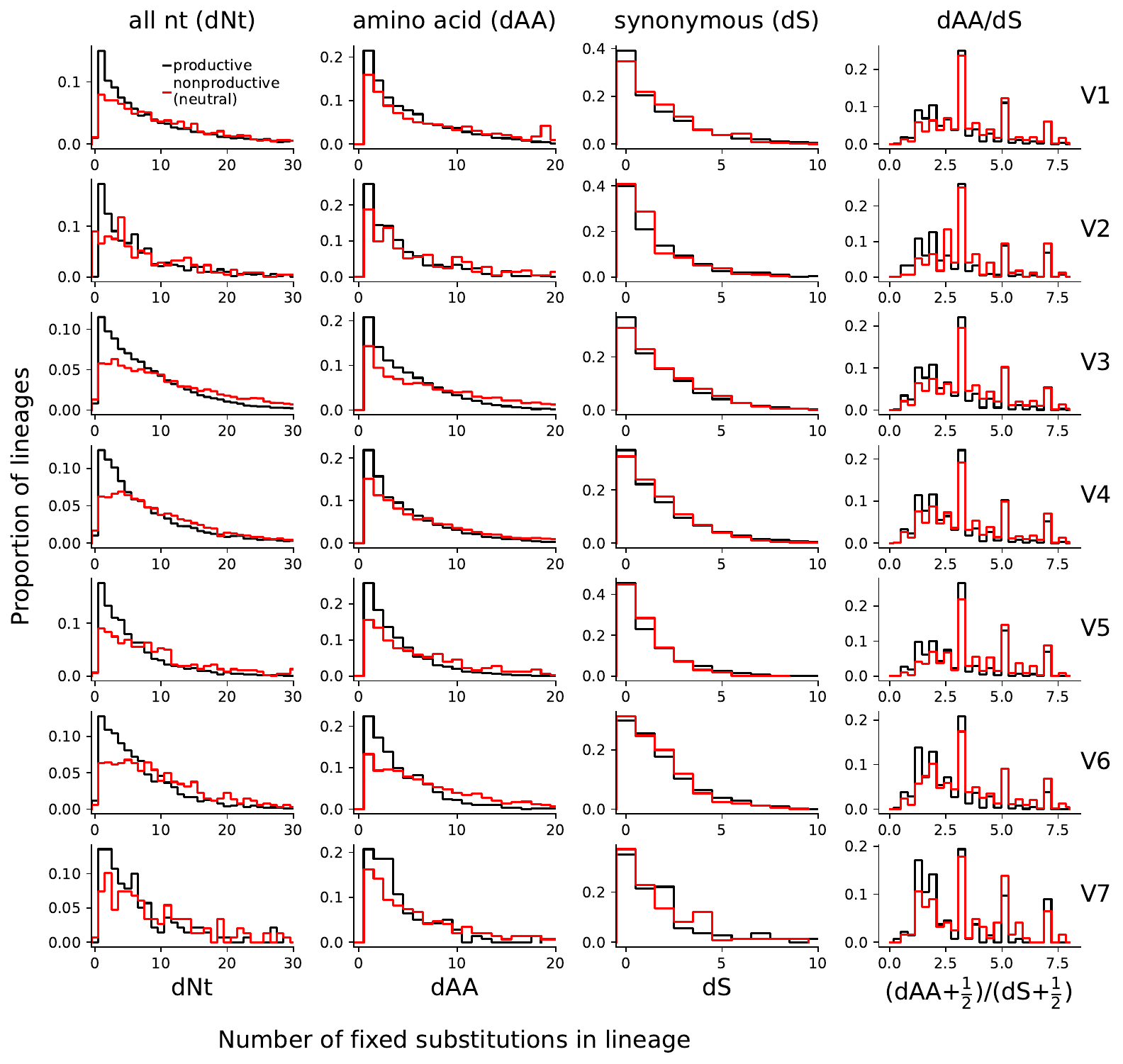}
    \caption{ \textbf{Similar distributions of fixed mutations in productive and nonproductive (neutral) lineages.} Distribution of fixed mutations in the V and J regions excluding the CDR3 (${\geq}80\%$ frequency among ${\geq}2$ sequences), among productive and nonproductive (putatively neutral) lineages~\siref{si:sequence processing}. For each V family (row), columns 1-3 correspond to, from left to right: nucleotide changes, amino acid changes, and synonymous nucleotide changes. Only lineages with at least one fixed amino acid change region are included; hence, the minimum divergence occupied in the middle plot is one.
    Column 4 presents the ratio of columns 2 to 3, analogous to an unnormalized $dN/dS$. Spikes in the distributions at $3,5,$ and $7$ reflect dAA${=}1,2,3$ and dS${=}0$. The $dAA/dS$ of productive affinity matured lineages is not biased to larger values than putatively neutrally evolving nonproductive lineages. This reflects the relatively weak signal of positive selection carried by $dN/dS$ in the context of affinity maturation.}
    \label{sfig:unproductive fixed divergence}
\end{figure}

\clearpage
\begin{figure}
    \centering
    \includegraphics[width=0.4\linewidth]{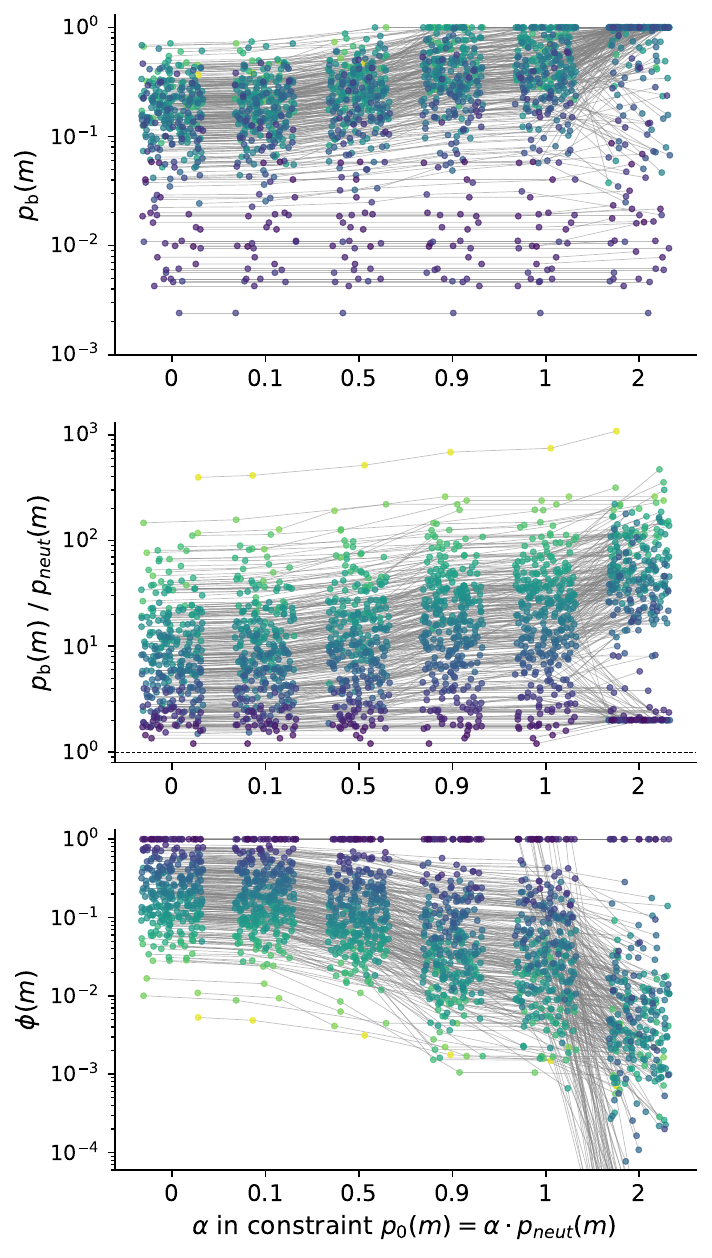}
    \caption{ \textbf{Mixture model inference is robust to choice of constraint.} The mixture model was re-inferred for alternative constraints $p_0 = \alpha p_{\rm neut}$, where $p_{\rm neut}$ was measured from out-of-frame nonproductive sequences~(\siref{si: mixture model}). Fig.~\ref{fig4:mixture model inference} corresponds to $\alpha=1$. Substitutions are shaded according to increasing inferred $p_{\rm b}/p_{\rm neut}$ at $\alpha=1$. Model inferences of the same substitution are connected by gray lines. Broadly consistent inferences across model constraints, particularly at large $p_{\rm b}$ (small $\phi$), reflect the sufficient separability of the beneficial and neutral classes.}
    \label{sfig:mixture model vs p0}
\end{figure}

\clearpage
\begin{figure}
    \centering
    \includegraphics[width=0.9\linewidth]{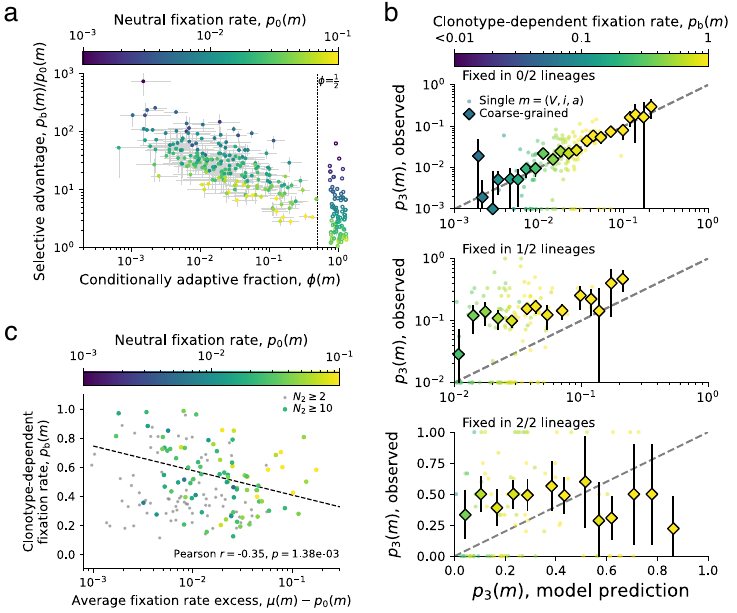}
    \caption{ \textbf{Inference meaningfully distinguishes $p_{b}/p_{0}$ from $\phi$.} (\textbf{a}) Analogue of Fig.~\ref{fig4:mixture model inference}b, normalizing $p_{\rm b}$ by $p_{0}$. As for the raw beneficial fixation rate, the ratio $p_{b}/p_{0}$ also anticorrelates with $\phi$. (\textbf{b}) Analogue of Fig.~\ref{fig4:mixture model inference}d-f, but instead of predicting with substitution-specific parameters, each substitution was assumed to have the same parameter, \DW{$p_{\rm b}/p_{0}=30$ and $\phi=0.02$, roughly corresponding to the central inference in \textbf{a} (note only mutations with $\phi < 1$ enter the prediction). This simplified model validates poorly, implying the parameter variation observed across substitutions in panel \textbf{a} and in Fig.~\ref{fig4:mixture model inference}b represents real biological variation. (\textbf{c}) Plotting $p_{\rm b}$ against the average fixation rate excess tests for a negative association between the frequency with which a mutation $m$ is beneficial across clonotypes (observed by the excess fixation $\mu-p_0$) and its conditional strength of selection (observed by $p_{\rm b}$), separate from the artificial tradeoff in Fig.~\ref{fig4:mixture model inference}b induced by the nature of the simplified mixture model~(\siref{si: mixture model}). Only mutations with a finite $\phi < 1$ are shown. Restricting to the best measured mutations (with 10 or more observed coincident fixations across eligible lineage pairs) reveals a modest but significant anticorrelation consistent with an empirical tradeoff.}
    }
    \label{sfig:pben/pneut vs pneut vs mode}
\end{figure}

\clearpage
\begin{figure}
    \centering
    \includegraphics[width=0.8\linewidth]{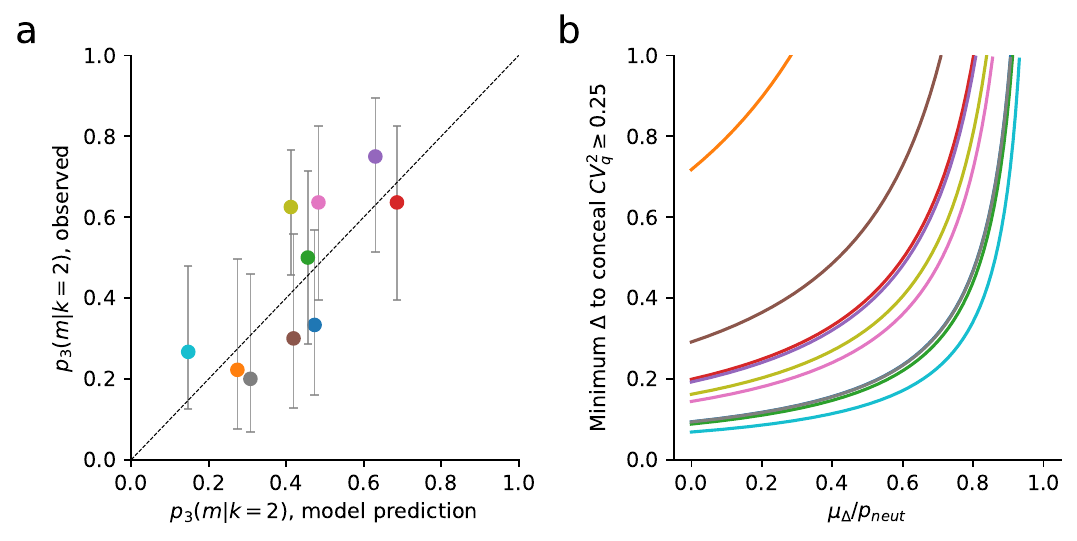}
    \caption{ \DW{\textbf{Substantial deleterious variation is necessary to mask beneficial variation.} (\textbf{a}) Analogue of Fig.~\ref{fig4:mixture model inference}f, restricted to the 10 best measured mutations $m=(V,i,a)$, namely those with the largest number $\geq 9$ of $k=2$-conditioned triplets to observe. Error bars represent 90\% Wilson confidence intervals. The consistent agreement between model prediction and observation (up to measurement error) implies that unmodeled variation in the beneficial class $\rho_{\rm b}(m)$ of support $p>p_0$ must be balanced, for each mutation, by the presence of unmodeled deleterious class of support $p<p_0$ (\siref{si: model interpretation}). (\textbf{b}) for each mutation in \textbf{a}, the scale of the deleterious class (summarized by its total weight $\Delta$ and average deleterious fixation rate $\mu_\Delta\equiv \langle p\rangle_\Delta < p_0$) required to conceal substantial beneficial class variation (summarized by a $\mathrm{CV}^2_q > 0.25$). To wit, most mutations would require lethal classes $p=0$ of size $\Delta \gtrsim 0.1$, or moderately deleterious classes $p \simeq 0.9 p_0$ of size $\Delta \rightarrow 1$, to mask $\mathrm{CV}^2_q > 0.25$. Note that these estimates do not depend on any 2-point model fits, nor on the specific shape of the deleterious class.}}
    \label{sfig:concealment of beneficial}
\end{figure}

\clearpage
\begin{figure}
    \centering
    \includegraphics[width=0.4\linewidth]{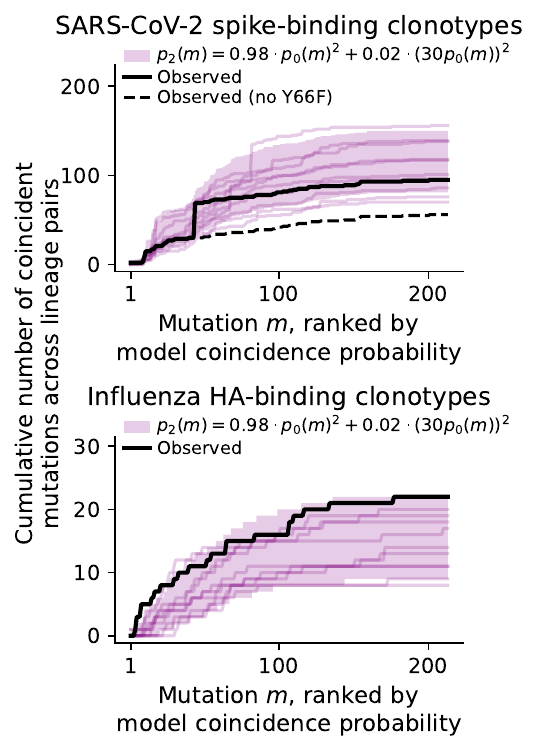}
    \caption{ \textbf{Analogue of Fig.~\ref{fig5:pathogen-specific antibodies}c, using a coarse-grained model inference}. Coincident fixation simulations among pathogen-specific clonotypes were re-simulated according to the clonotype-specific model of coincidence.  In this case, mutation-specific $\{\phi(m),p_{\rm b}(m)\}$ were replaced with uniform \DW{$\phi(m)=0.02$, $p_{\rm b}(m)=30\cdot p_0(m)$}, as in Fig.~\ref{sfig:pben/pneut vs pneut vs mode}. We find comparable agreement to the per-mutation inferred $\{\phi(m), p_{\rm b}(m)\}$, reflecting the fact that the small number of pathogen-specific clonotypes considered here provide insufficient statistics to test averaged versus local features of the clonotype-dependent adaptive landscape.
    }
    \label{sfig: antigen vs coarsegrain}
\end{figure}

\begin{figure}
    \centering
    \includegraphics[width=1.0\linewidth]{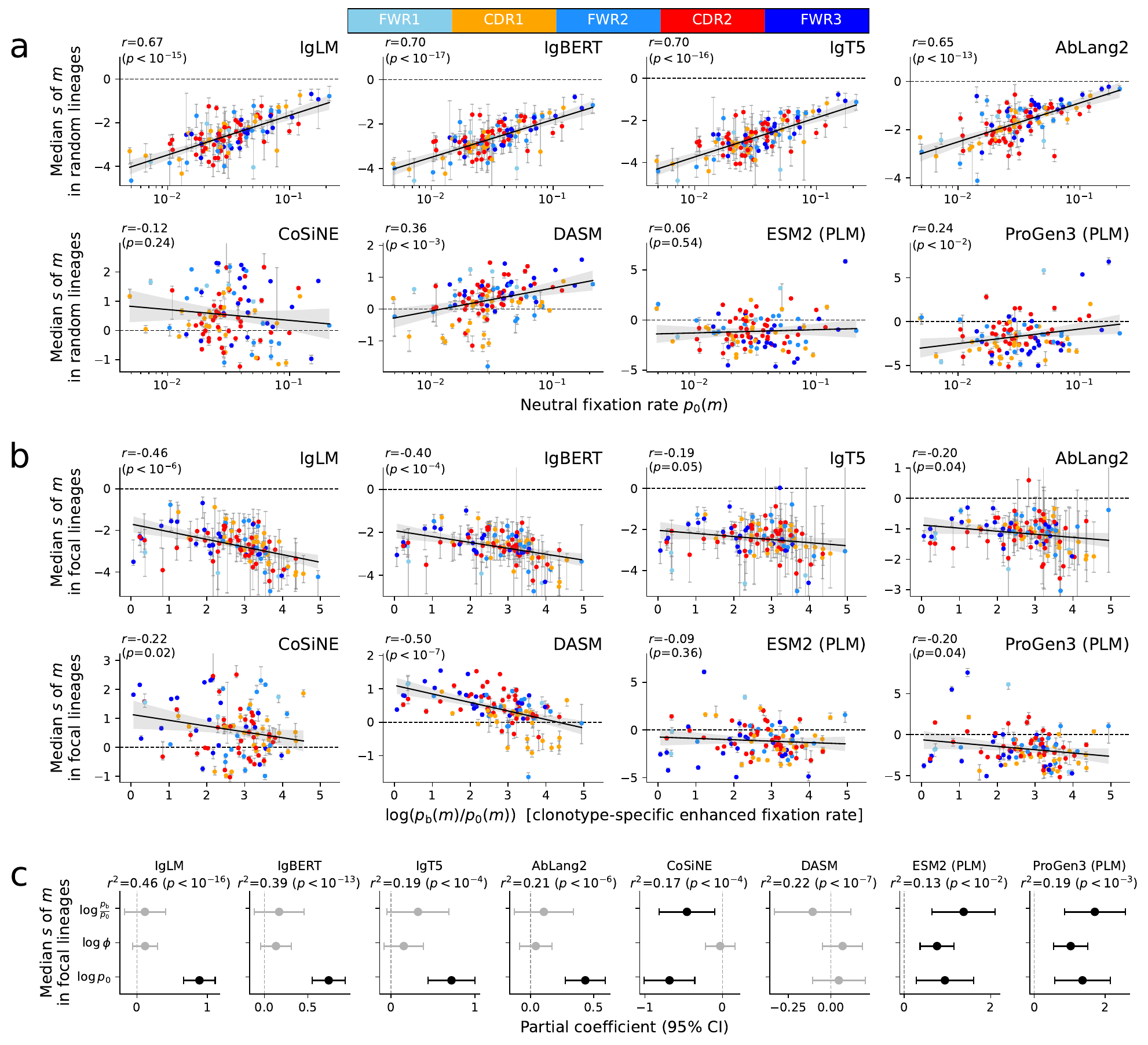}
    \caption{ \DW{\textbf{Antibody LM scores correlate well with clonotype-independent repertoire features but poorly with clonotype-specific substitution fitness measures.} (\textbf{a}) Across 110 mutations (${\geq}10$ coincident fixations each in lineage pairs of shared clonotype), median score in randomly sampled lineages correlates with neutral fixation rate $p_0$ in  standard ALMs (top row), and more weakly or not at all in SHM-corrected ALMs or SHM-unaware protein language models (bottom row). (\textbf{b}) Across same mutations, median score in focal lineages anticorrelates with the clonotype-specific enhanced fixation rate across antibody language models. (\textbf{c}) Across 151 mutations (${\geq}5$ coincident fixations in lineage pairs of shared clonotype), multiple linear regression with intercept of the median score in focal lineages against $\log p_0$, $\log\phi$, and $\log p_{\rm b}/p_0$ (Methods). Each LM panel reports a coefficient of determination $r^2$ and $p$-value against an intercept-only model (F-test), as well as the partial coefficients for each feature (95\% confidence intervals). Across most ALMs, $\log p_0$ carries the significantly non-zero partial coefficient. Curiously, each feature has a significant partial coefficient for protein language model scores, albeit with only a modest overall variance explained.}
    }
    \label{sfig: LM score vs pneut}
\end{figure}

\begin{figure}
    \centering
    \includegraphics[width=1.0\linewidth]{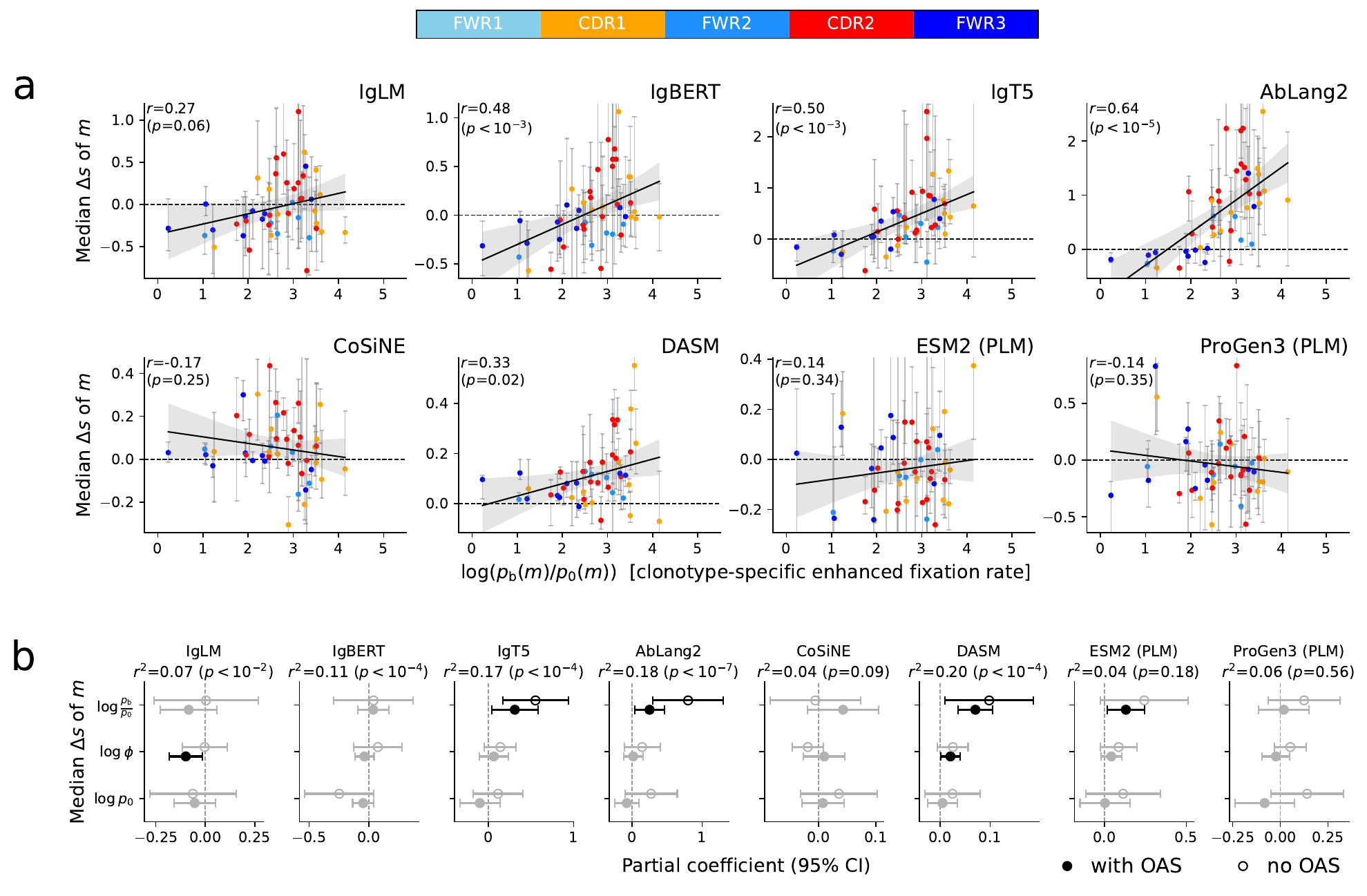}
    \caption{ \textbf{Analogue of Fig.~\ref{fig6:language models}c, holding out clonotypes sourced from OAS subjects (B19 and P22).} (\textbf{a}) The persistent signal in \DW{IgBERT, IgT5, and AbLang2, which were trained on OAS subject repertoires, as well as DASM, which was not, suggests that these models have not simply memorized mutations to produce the correlations in Fig.~\ref{fig6:language models}c. Only 47 mutations with 10 or more coincidences among non-OAS pairs were included. (\textbf{b}) Multiple regression of the median $\Delta s$, analogous to Fig.~\ref{sfig: LM score vs pneut}c. For IgT5, AbLang2, and DASM, positive partial coefficients are consistently fit to $\log p_{\rm b}/p_0$, implying these ALMs capture some general patterns of clonotype-specific enhanced fixation rates. These results hold for both OAS ($n=151$ mutations) and OAS-excluded ($n=83$) data.}}
    \label{sfig: LM deltasrcore no OAS}
\end{figure}

\end{document}